\documentclass[twocolumn]{aastex631}
\usepackage{amsmath}    

\newcommand{\new}[1]{{\textcolor{black} {#1}}}
\newcommand{\llrw}[1]{{\textcolor{black} {#1}}}

\begin{document}

\title{Flashlights: Properties of Highly Magnified Images Near Cluster Critical Curves in the Presence of Dark Matter Subhalos}

\author[0000-0002-6039-8706]{Liliya L.R. Williams}
\affiliation{Minnesota Institute for Astrophysics, University of Minnesota, 116 Church Street SE, Minneapolis, MN 55455, USA}

\author[0000-0002-0786-7307]{Patrick L. Kelly}
\affiliation{Minnesota Institute for Astrophysics, University of Minnesota, 116 Church Street SE, Minneapolis, MN 55455, USA}

\author[0000-0002-8460-0390]{Tommaso Treu}
\affiliation{Department of Physics and Astronomy, University of California, Los Angeles, CA 90095}

\author[0000-0003-1276-1248]{Alfred Amruth}
\affiliation{Department of Physics, University of Hong Kong, Pokfulam Road, Hong Kong}

\author[0000-0001-9065-3926]{Jose M. Diego}
\affiliation{IFCA, Instituto de F\'isica de Cantabria (UC-CSIC), Av. de Los Castros s/n, 39005 Santander, Spain}

\author[0000-0002-4490-7304]{Sung Kei Li}   
\affiliation{Department of Physics, The University of Hong Kong, Pokfulam Road, Hong Kong}

\author[0000-0002-7876-4321]{Ashish K. Meena}
\affiliation{Physics Department, Ben-Gurion University of the Negev, P.O. Box 653, Beer-Sheva 8410501, Israel}

\author[0000-0002-0350-4488]{Adi Zitrin}
\affiliation{Physics Department, Ben-Gurion University of the Negev, P.O. Box 653, Beer-Sheva 8410501, Israel}

\author[0000-0002-8785-8979]{Thomas J. Broadhurst}
\affiliation{Department of Physics, University of the Basque Country UPV/EHU, E-48080 Bilbao, Spain}

\author[0000-0003-3460-0103]{Alexei V. Filippenko}
\affiliation{Department of Astronomy, University of California, Berkeley, CA 94720-3411, USA}

\begin{abstract}
  Dark matter subhalos with extended profiles and density cores, and globular star clusters of mass $10^6$--$10^8\,M_\odot$, that live near the critical curves in galaxy cluster lenses can potentially be detected through their lensing magnification of stars in background galaxies. In this work we study the effect such subhalos have on lensed images, and compare to the case of more well-studied microlensing by stars and black holes near critical curves. We find that the cluster density gradient and the extended mass distribution of subhalos are important in determining image properties. Both lead to an asymmetry between the image properties on the positive and negative parity sides of the cluster that is more pronounced than in the case of microlensing. For example, on the negative parity side, subhalos with cores larger than about $50\,$pc do not generate any images with magnification above $\sim 100$ outside of the immediate vicinity of the cluster critical curve. We discuss these factors using analytical and numerical analysis, and exploit them to identify observable signatures of subhalos: \new{subhalos create pixel-to-pixel flux variations of $\gtrsim 0.1$\,mag, on the positive parity side of clusters. These pixels tend to cluster around (otherwise invisible) subhalos.}  Unlike in the case of microlensing, signatures of subhalo lensing can be found up to $1''$ away from the critical curves of massive clusters. 
\end{abstract}

\section{Introduction}

Recent work has shown that high magnification near galaxy cluster critical curves makes it possible to detect individual stars in background, strongly lensed galaxies at intermediate and high redshifts \citep{kel18,che19,kau19,che22,kel22,wel22,mee22a,mee23}, realizing a prediction made three decades ago \citep{mir91}. The recent 192-orbit {\it Hubble Space Telescope (HST)} Flashlights project \citep[PI P. Kelly, GO-15936;][]{kel22} was specifically designed to look for such high-magnification events in the six Hubble Frontier Field galaxy clusters \citep{lot17}.
Lensing theory has evolved rapidly in the last several years to explain and take advantage of these new exciting events \citep{ven17,dai21,die22,mee22b}.

Theory predicts that on the two sides of the cluster critical curve, certain image properties are different. On the side farther away from the cluster center, images have the same handedness as the unlensed (unobservable) sources, while on the side closer to the cluster center images are mirror-imaged compared with the source. The two sides are said to have positive and negative parities, respectively, and the corresponding images are called minima and saddle points, referring to the type of extrema they represent in the arrival time, or Fermat surface \citep{bla86,sch92}. While parity cannot be observed for images of point sources like stars, studies of microlensing by intracluster stars \llrw{\citep{die18,ogu18,kau19}}, and possibly black holes near cluster critical curves, show that magnification properties of lensed images on the positive and negative parity sides of the cluster are different: for the former, the minimum magnification is within a factor of a few of the smooth cluster's magnification at that location, and can be as high as thousands, whereas the magnification of the latter can be smaller by factors of $\sim 10$--100, and so these images can remain invisible for long stretches of time \citep[see Figures 6 and 8 of][]{die18}. This property is related to the lensing theorem that minima can never be demagnified, while there is no such restriction on saddles \citep{sch92}. 

High magnification of these events not only helps with the study of otherwise undetectable stars, but also allows for better resolution of the lensing mass  distribution in the galaxy cluster, including intracluster stars and dark matter subhalos.  Lensing by subhalos near cluster critical curves is an emerging topic \citep{dai18,dai20}, and is also the subject of this paper. Dark matter subhalos share some similarities with stars and black holes as lenses; however, there are also important differences. 

\llrw{In this paper we study the effect of subhalos with finite central densities and hence core radii, that live in clusters whose density falls with distance from the center. We focus on the asymmetry between the positive and negative parity sides of the cluster, and use the asymmetry to devise tests to detect subhalos, and possibly determine the mass fraction and properties of subhalos, namely the core radius of their mass distribution. We obtain some analytical approximate results, emphasize the role of cluster density gradient and subhalo core radii, and discuss the similarities and differences between subhalo lensing and microlensing by stars.
We work with lensing simulations that solve the lens equation to find images of sources, leaving the application to observations to a later paper.} \\

\section{The Lens Mass Model}

The lens model we use for our simulations consists of the cluster macromodel, and a zoom-in portion of the cluster, near its tangential critical curve, populated with subhalos. 

\subsection{Cluster Model}

The cluster is smooth and elliptical, with a small core and isothermal density profile, $\propto {r_{2D}}^{-1}$. In Appendix~\ref{appB} we present results that use a somewhat shallower cluster density profile, $\propto {r_{2D}}^{-0.9}$.  Such slope values are typical in the regions where multiple images are found \citep[e.g., ][]{jau19,gho22}.
Its critical curve is shown as the black curve in Figure~\ref{fig:images00}; here we will refer to it as the cluster CC. To model the effects of subhalos we use a square modeling window (gray), which encloses two merging images (red circles) of a source depicted by a blue star. The other images of that source are also shown as red circles, but we do not use these in the rest of the paper. We populate this square with subhalos, as described later, and leave the rest of the cluster subhalo-free.

We scale our cluster model to the observed SGAS J1226+2152 cluster, at redshift $z_l=0.43$, used by \cite{dai20}. The image pair near the cluster CC that the authors analyzed originates from a source at $z_s=2.93$. For standard $\Lambda$CDM cosmology, that implies a critical surface mass density for lensing of $\Sigma_{\rm crit}=0.41$\,g\,cm$^{-2}$, and $1''$ is 5.61\,kpc in the lens plane. The distance to the merging pair from the cluster center is $7.3''$. Using this scaling, the size of the modeling frame (gray square) is $2.0''$ across, or 11.2\,kpc in the lens plane. The total projected cluster mass in the modeling window is $1.243\times 10^{11}\,M_\odot$. We note that SGAS J1226+2152 has a rather low mass, so when comparing our sky-projected distances quoted in arcsec to that of some other cluster, one needs to rescale the distances. For example, MACS 1149 has an Einstein radius of $\sim 30''$ for sources at a redshift similar to the one used here, a factor of $\sim 4.2$ larger. Other massive clusters could have even larger sizes \citep[see Table 5 of ][]{joh14}.

For the purposes of calculations, our code length unit in the lens plane \llrw{(i.e., the resolution used to initially calculate image positions)} is $0.002''$. To calculate \llrw{final} image positions and magnifications we zoom in by a further factor of 10, to $2\times 10^{-4}$ arcsec. In the analysis \llrw{that calls for} the observable properties, we use coarser pixels, $0.04''$, that are closer \llrw{to the typical pixel scale usually used for {\it HST} reductions such as the Frontier Field\footnote{\llrw{Data reduction for Frontier Fields data used stacked and drizzled images at scales of $0.03''$ and $0.06''$.}} \citep{lot17}}.

\subsection{Sources and Their Images}\label{sec:source}

\begin{figure}
    \centering
    \vspace{-2cm}
    \includegraphics[width=0.45\textwidth]{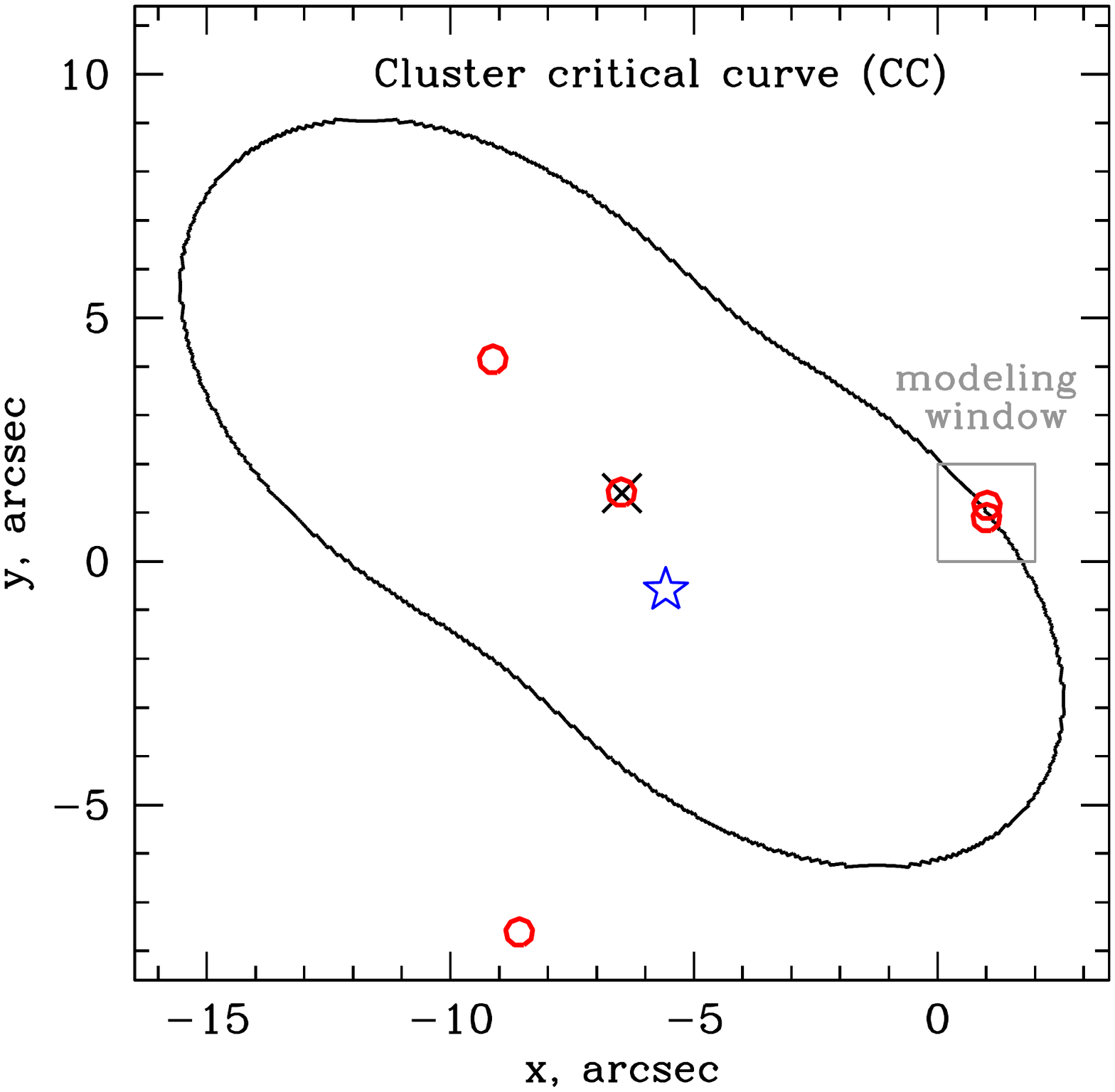} 
    \vskip-2cm
    \caption{Cluster-scale critical curve. The source is the blue star symbol and its images are shown with red circles. The merging pair of images straddling the CC that we will model in detail is on the right side of the cluster. The box around it is the size of the modeling window we use, for example in  Figure~\ref{fig:smmass}. The lower-left corner of the box defines the origin of our coordinates. Our Cartesian coordinate system is oriented such that the image stretching is very nearly along the vertical $y$ axis, which aides in the analysis in Section~\ref{sec:reasons}. \\ }
    \label{fig:images00}  
\end{figure}

Our source is a portion of a galaxy, $\sim 2'' \times 0.5''$, near the blue star symbol in Figure~\ref{fig:images00}, such that it fills the modeling window in the lens plane. It consists of individual stars, represented by point sources, \new{but our numerical calculations will hold for any source no larger than $\sim 1\,$pc.}\footnote{\new{Assuming image magnification $\propto$ (subhalo caustic size/source size)$^2$, and given that the highest magnification in our calculations is $\sim 10^4$ and the caustic size for the typical subhalo is $\sim 0.01''$, the source cannot be larger than $\sim {10^{-4}}''$, or $\sim 1\,$pc, to be modeled as a point source.}} The sources are randomly scattered in the source plane, \new{but one could also select a subset of these that would follow a correlation function, such as the one found for nearby star-forming regions}. The average surface density of stars in the source plane is 0.12\,stars\,pc$^{-2}$. This density is lower than that of all stars in an $L_\star$ galaxy like the Milky Way. However, only the intrinsically brightest stars, like giants, are interesting in the present context, and their number density is lower, especially if the search is done in a specific filter that would be sensitive to stars of a given type.

We solve the lens equation, forward lensing a few $\times 10^6$ stars per realization, but only about a third of them end up generating at least one image in the lens-plane window. Image multiplicity ranges from 1 to 3--4 within the lens-plane modeling window, but the typical multiplicity is 1 or 2, and the typical number of images is $2\times 10^6$.

In some of our calculations we are interested in lensing magnification, so these calculations assume that all stars have the same luminosity. Later in the paper, when considering observables, we use a luminosity function for stars, $dn/dL\propto L^{-2.5}$ \citep{kel18}. The luminosity normalization is not important because it merely rescales all fluxes by the same multiplicative factor, while our analysis deals with flux ratios, or lensing magnifications.

\subsection{Subhalos in the Lens Plane}\label{sec:lens}

\begin{figure}
    \centering
    \includegraphics[width=0.45\textwidth]{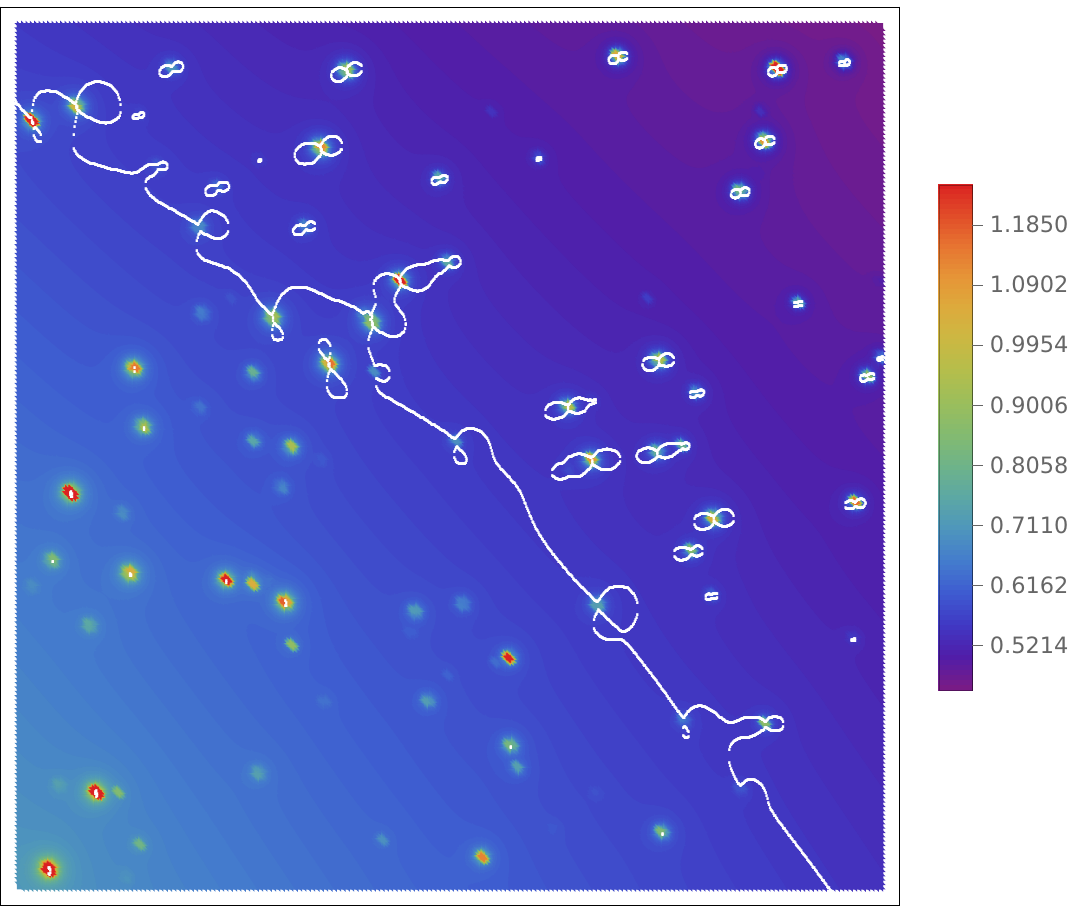}
    \caption{An example of the lens-plane modeling window (gray box in Figure~\ref{fig:images00}) populated with subhalos whose mass fraction within this window is 1\%. The color shows the $\kappa$ density. The frame is $2''$ on a side, the white line is the critical curve, and the center of the cluster is to the left of the frame.\\}
    \label{fig:smmass}
\end{figure}

We populate the small gray square in Figure~\ref{fig:images00} with subhalos. One example of the total projected density is shown in Figure~\ref{fig:smmass}. The $\kappa$ gradient due to the main cluster is clearly visible, while subhalos generate small deviations from a smooth density gradient.

Subhalos were drawn from a mass function, $dn/dm\propto m^{-0.9}$, used by \cite{dai20}. They cover two decades in mass, $10^6 M_\odot\rightarrow 10^8 M_\odot$. With this, the average mass of subhalos is $\langle m\rangle=2.46\times 10^7\, M_\odot.$ Guided by previous work \citep{gil19,dai20}, we use mass fractions in subhalos of about 1\% and 0.5\%, and in some cases 0.25\%. 
With 1\% (0.5\%) the number of subhalos in the modeling window of $2'' \times 2''$ is $\sim 86$ ($\sim 43$). Note that because of the finite number of subhalos, and the stochastic nature of random draws from the mass function, one cannot guarantee that any given realization of the lens-plane window will have exactly 1\% or 0.5\% mass in subhalos. The actual mass fractions in subhalos in our runs tend to be somewhat different from these values, and are indicated in the relevant figures.

If subhalos were added over the whole extent of the cluster, then the surface mass density of the smooth component of the cluster would need to be reduced by the corresponding amount, to keep the total density the same. However, we add subhalos only to the modeling window (small gray square in Figure~\ref{fig:images00}); adding subhalo mass fraction of 1\% translates into increasing the density (and hence the deflecting mass of the cluster) by $\sim 0.01$\%, which is negligible. This is further supported by the fact that the distance of the critical curve from the cluster center is not affected by the addition of subhalos. 

The density profile of subhalos is given by, for example, \ref{eq:kappa} in Appendix~\ref{appA}. For $a=0.05$ it closely resembles the Navarro-Frenk-White profile for about three decades in radius, but falls off more sharply at large radii, imitating tidal truncation expected for subhalos living in a dense cluster environment. The simple analytical form of its relevant lensing properties aides greatly in reducing computational cost. 

We take the core radius of the model to correspond to the radius where the three-dimensional (3D) slope is approximately ``isothermal,'' at $r\approx 0.1$ in the natural units of the profile; see Figure~\ref{fig:myy}. This core radius needs to correspond to a certain physical size that we estimate from published data. Consulting the LITTLE THINGS galaxy survey \cite[Table 2 of][]{ohx15}, the typical radii of $10^9\, M_\odot$ galaxies is about 300\,pc, with substantial variation between individual galaxies. This value is consistent with the results of \cite{wol10}. Our subhalos are less massive, so these values have to be scaled down. We chose the scaling motivated by the virial condition, where the cube of the virial radius scales linearly with the virial mass; we use $r_{\rm core}\propto m^{1/3}$, even though the core radius need not scale linearly with the virial radius. The average subhalo mass is $\sim 40$ times less massive than $10^9\, M_\odot$, so core radii should be a factor of $\sim 3.5$  smaller, or $\sim 85$\,pc. 

If our subhalos are to represent globular star clusters instead of dark matter subhalos, then their core radii should be smaller, of order a few tens of parsecs \citep{fai22}. To span the approximate range of possible core radii, we do three sets of calculations, with 22\,pc, 44\,pc, and 88\,pc, but show only the 22\,pc and 88\,pc results in most of the plots. The mass range of our subhalos, $10^6$--$10^8\, M_\odot$, is reasonably appropriate for the higher mass end of globular clusters.

Globular clusters will have a similar gravitational lensing effect as subhalos, but globulars would contribute to the observable light, thereby altering the distribution of flux in the lens plane. The number density of globulars in the strong lensing regions of clusters is $\sim 1$ per square arcsec \citep{lee22,die23}, so our modeling window would have $\sim 4$ globulars, considerably fewer than dark matter subhalos \citep{gil17,mao98}.  Therefore, we do not consider globulars as subhalos in this paper; we assume that all subhalos are completely dark.

For each run, a few $\times 10^6$ stars from the extended source (Section~\ref{sec:source}) were forward lensed. All were multiply imaged by the cluster as whole, and some were further multiply imaged by subhalos. The fraction of sources that were multiply imaged by subhalos depends on the density profile of subhalos. Massive subhalos with density cusps, or small core radii, can generate several additional images, but for the subhalo core radii chosen here, image multiplicity above 2 is rare. Lensed images that ended up outside of the lens plane modeling window were ignored.  

One realization of the modeling window is not sufficient to draw statistically meaningful conclusions, because the typical number of subhalos in any 2 square arcsec region is small, $\sim 20$--80, and given that subhalos are drawn from a mass function and their positions are randomly chosen results in considerable scatter between realizations. Therefore, we generate 16 realizations of each subhalo scenario. Most figures in this paper that show statistical results use 16 realizations.

\section{The Effect of Subhalos on the Distribution and Number Density of Highly Magnified Images}

In this section, we present the effect that subhalos have on the image magnification and distribution in the lens plane. We will be using magnifications, not fluxes, which is equivalent to assuming that all sources have the same luminosity. While magnifications and exact positions of individual images are not directly observable, understanding them will aid in devising observable tests for subhalos.

First, a few notes about the abbreviated names we use for certain commonly occurring concepts. We will refer to the positive and negative parity sides of the cluster CC simply as the +ve and $-$ve sides of the cluster. By cluster ``sides'' we mean the two portions of the cluster on either side of the CC of the subhaloless cluster. We use this same definition for both subhaloless and subhaloed clusters. In other words, the cluster CC always refers to that of the subhaloless cluster. CC of subhaloed cluster often has wiggles and loops, and it would be hard to differentiate between its two ``sides.'' Subhalos themselves, especially those farther from the cluster CC, often have their own CC loops, and we will specifically indicate that in the text. 

The +ve and $-$ve sides of the cluster are not to be confused with the +ve and $-$ve parities of {\it{images}}. While most of the images on the +ve cluster side have +ve parity, that is not always the case in the presence of subhalos. The same is true of the $-$ve cluster side and $-$ve parity images.

\begin{figure*}[!t]
    \centering
    \vspace{-2cm}
    \includegraphics[width=0.495\textwidth]{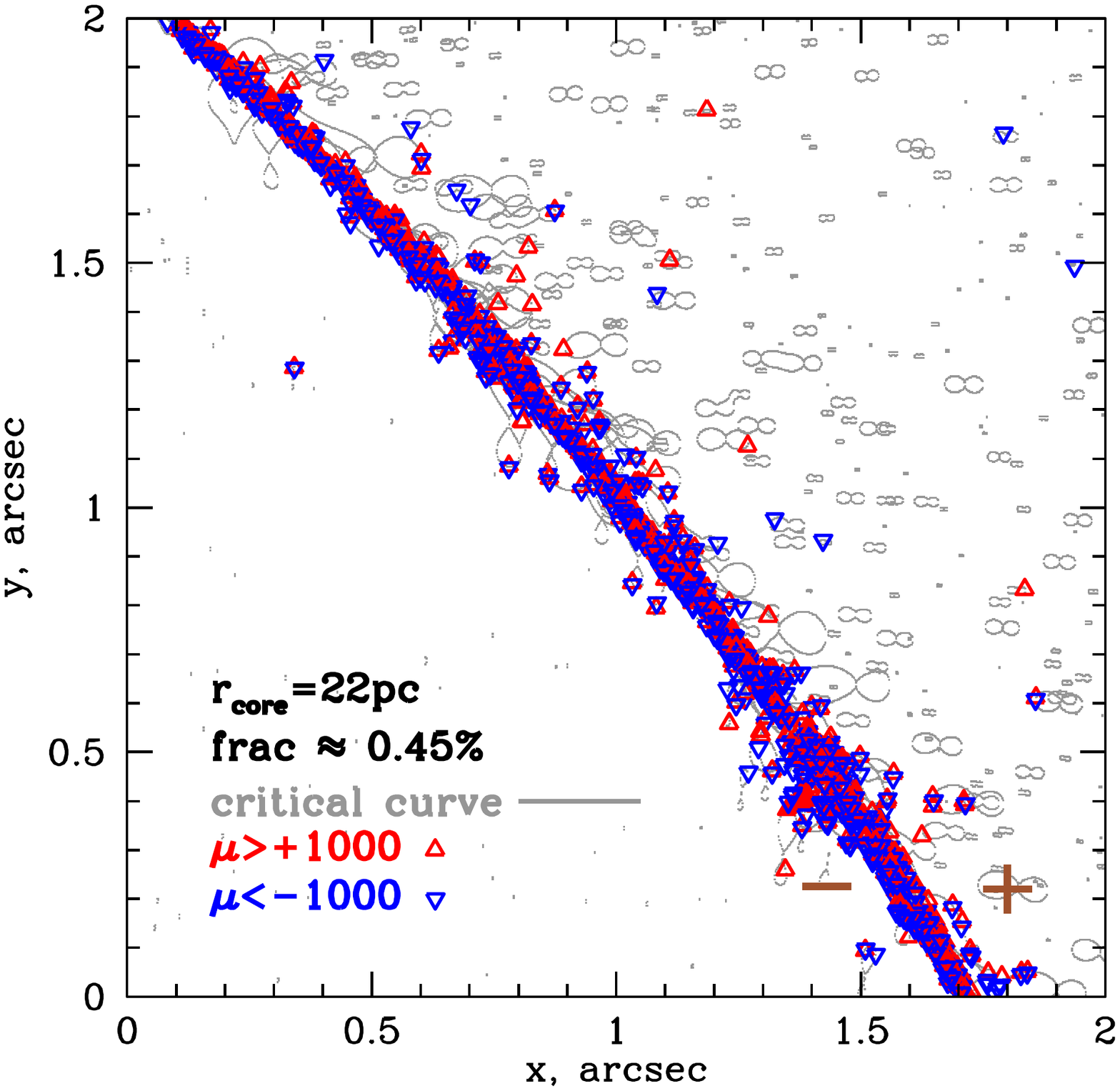}
    \includegraphics[width=0.495\textwidth]{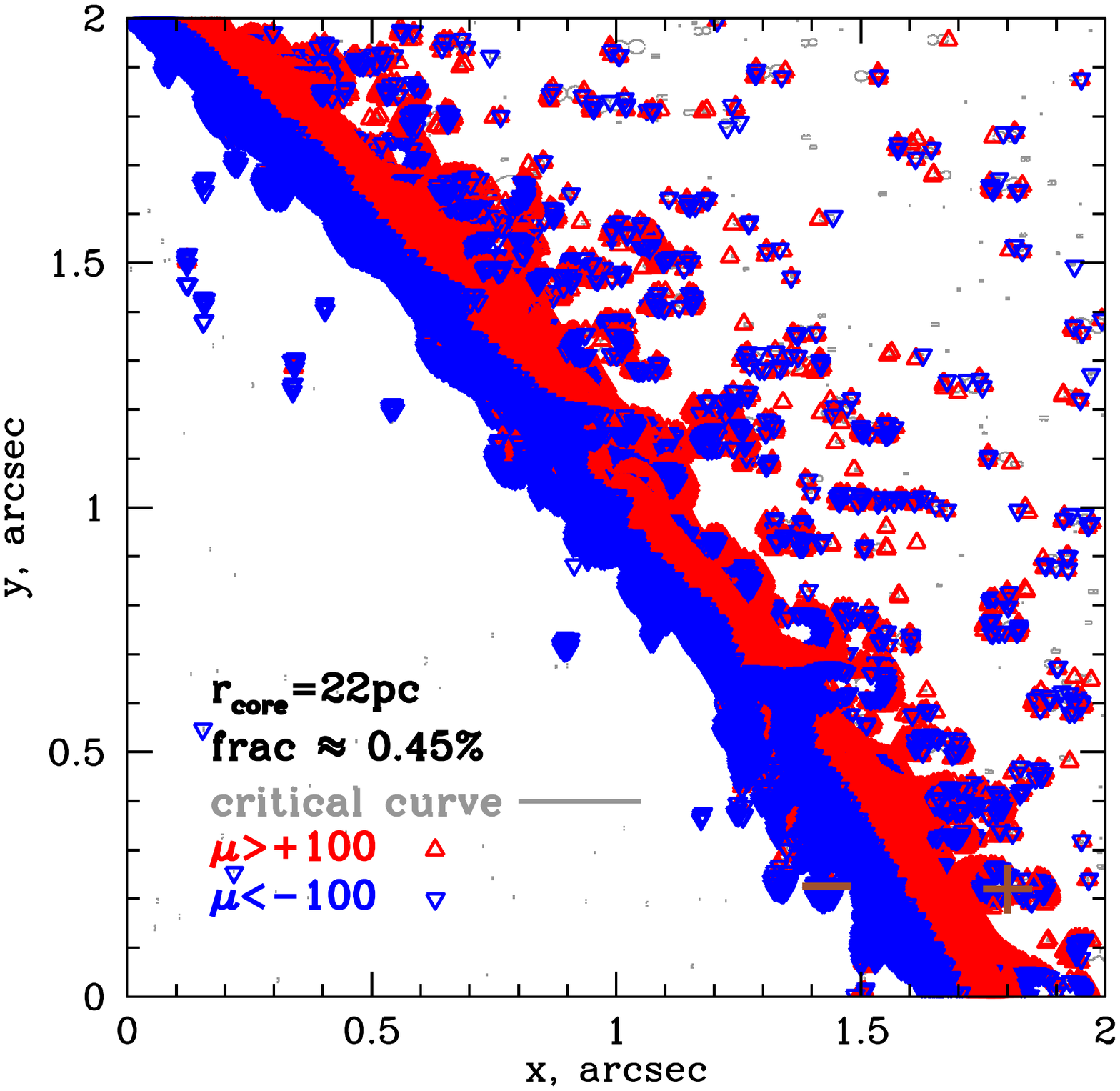} \vskip-4.4cm
    \includegraphics[width=0.495\textwidth]{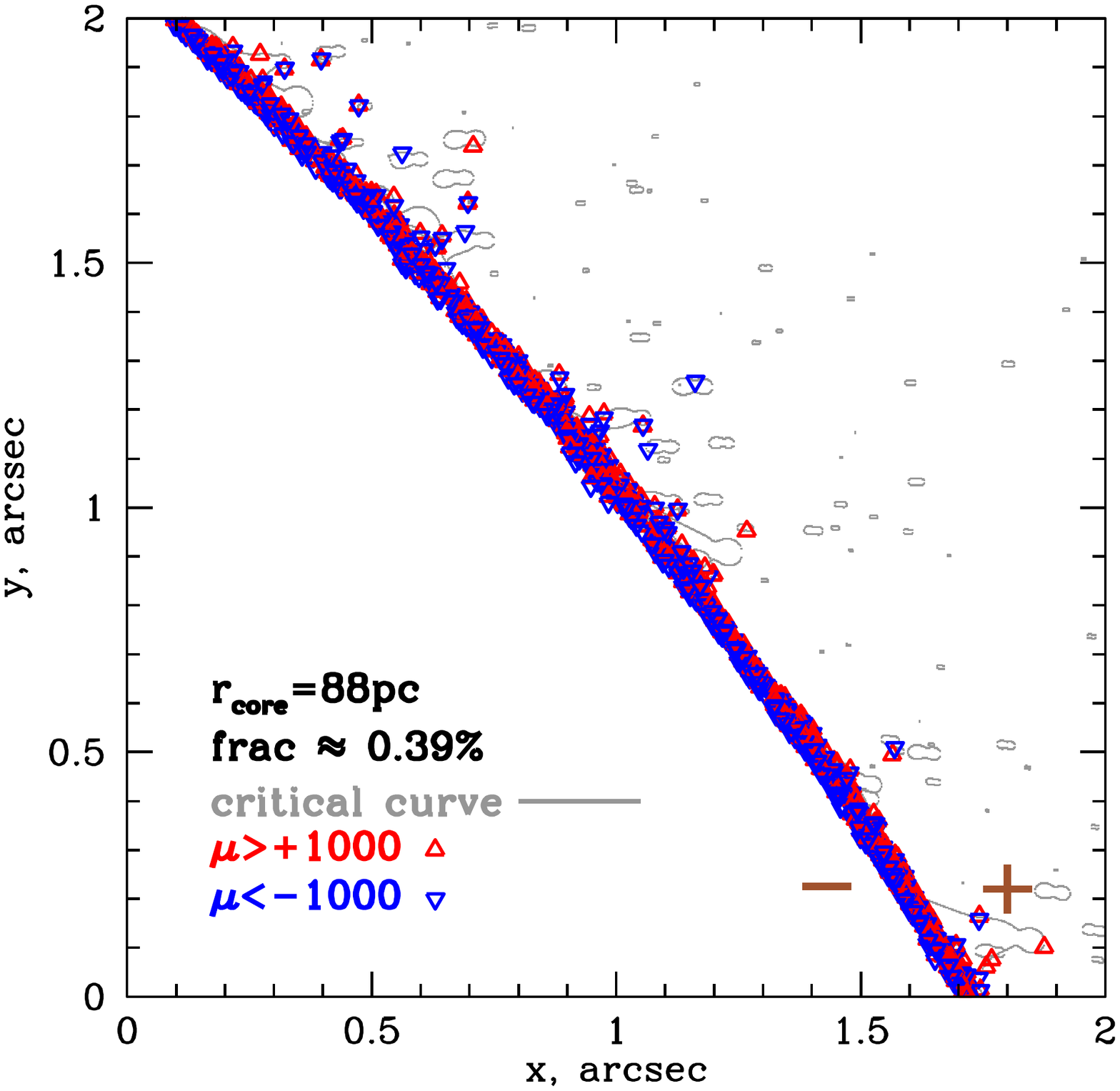}
    \includegraphics[width=0.495\textwidth]{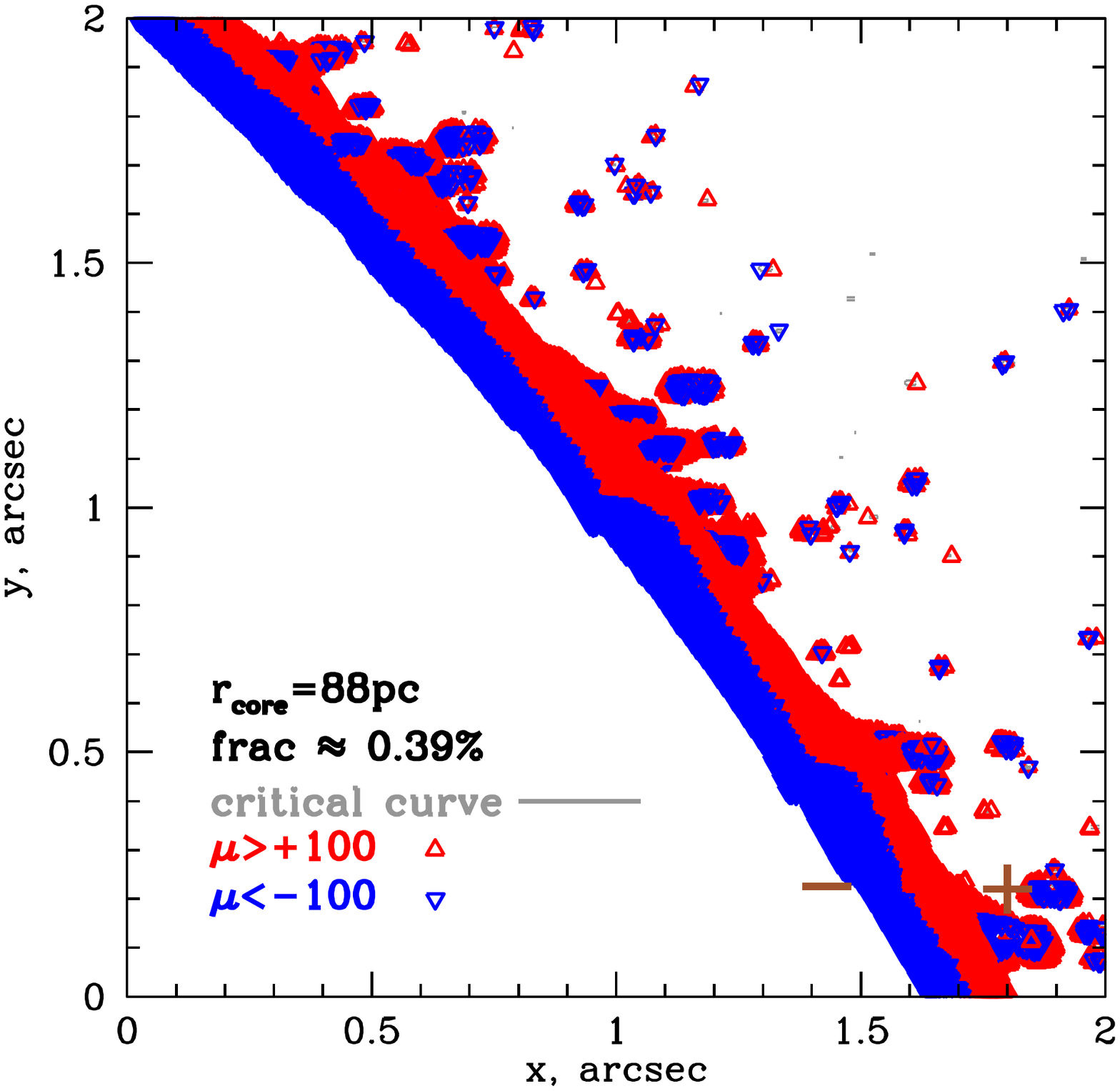} \vskip-2.2cm
    \caption{A $2\,\square ''$ box in the lens plane, showing a piece of cluster CC in the presence of subhalos with mass fraction $\sim 0.5\%$. The figures contain a superposition of 16 different realizations, to enhance the visualization of the effect of subhalos.  Subhalos themselves are not marked to avoid overcrowding. Critical curves are shown as gray. The +ve and $-$ve parity sides of the cluster are indicated with `+' and `-' brown signs.  The projected density in the cluster is $\rho\propto {r_{2D}}^{-1}$.  The number of subhalos per realization is $\sim 44$.
    {\it Upper Left:} Subhalo core radii are $r_c=22\,$pc. Images with +ve and $-$ve parity with $|\mu|>1000$ are red and blue, respectively. 
    {\it Upper Right:} Subhalo core radii are $r_c=22\,$pc. Images with +ve and $-$ve parity with $|\mu|>100$ are red and blue, respectively.    
    {\it Lower Left:} Subhalo core radii are $r_c=88\,$pc. Images with +ve and $-$ve parity with $|\mu|>1000$ are red and blue, respectively. 
    {\it Lower Right:} Subhalo core radii are $r_c=88\,$pc. Images with +ve and $-$ve parity with $|\mu|>100$ are red and blue, respectively.    
    Note that aside from the band immediately around the cluster CC, highly magnified images in all panels are found preferentially on the +ve parity side. See Figure~\ref{fig:countnearCC} for the histogram of image distribution as a function of distance from the CC.}
    \label{fig:imageplane}
    \vspace{0.25cm}
\end{figure*}

\subsection{Distribution of Bright Images in the Lens Plane}

The \llrw{four panels} of Figure~\ref{fig:imageplane} show a portion of the lens plane near the cluster CC (gray box in Figure~\ref{fig:images00}). To enhance the visual impression of the effect of subhalos, we show a superposition of 16 independent realizations, each with a subhalo mass fraction of $\sim 0.5\%$. The gray curves are the CC in the presence of subhalos. The brown ``+'' and ``-'' signs label the +ve and $-$ve sides of the cluster. The red and blue points are images of +ve and $-$ve parity, respectively, all with $|\mu|>1000$ (left panels), and with $|\mu|>100$ (right panels). \llrw{Lensing magnification means that the lens-plane area in the region where that magnification applies is stretched by the magnification factor. As a consequence,} the number density of \llrw{images, including} highly magnified images, is reduced. Bright images are present predominantly on the +ve side. For large subhalo core radii (lower panels), bright images are completely absent from the $-$ve side, if one excludes a band around the cluster CC. \llrw{As we discuss later, large subhalo core radii result in a reduction of the length of subhalo CC's and the number of bright images.}

Subhalos displace the positions of images from where they would have formed in a smooth cluster. The typical displacement is $0.01''$--$0.02''$, and would be undetectable, especially given lens mass model uncertainties \citep[see Figure 2 of][]{che19} and the image lens plane root-mean square (RMS; see Table 1 of \cite{pri17}, and, for example, \cite{ber23}).

As expected, highly magnified images outline the cluster CC, but the presence of subhalos makes the CC deviate from that of the subhaloless case. In some places the CC is so distorted that the string of highly magnified images is significantly away from the CC of subhaloless cluster, by $\sim 0.1''$, for a cluster of mass comparable to that of SGAS J1226+2152. This results in a larger number of highly magnified images.

\begin{figure}[!h]
    \centering
    \vspace{-2cm}
    \includegraphics[trim={0.2cm 0 0cm 0cm},clip,width=0.5\textwidth]{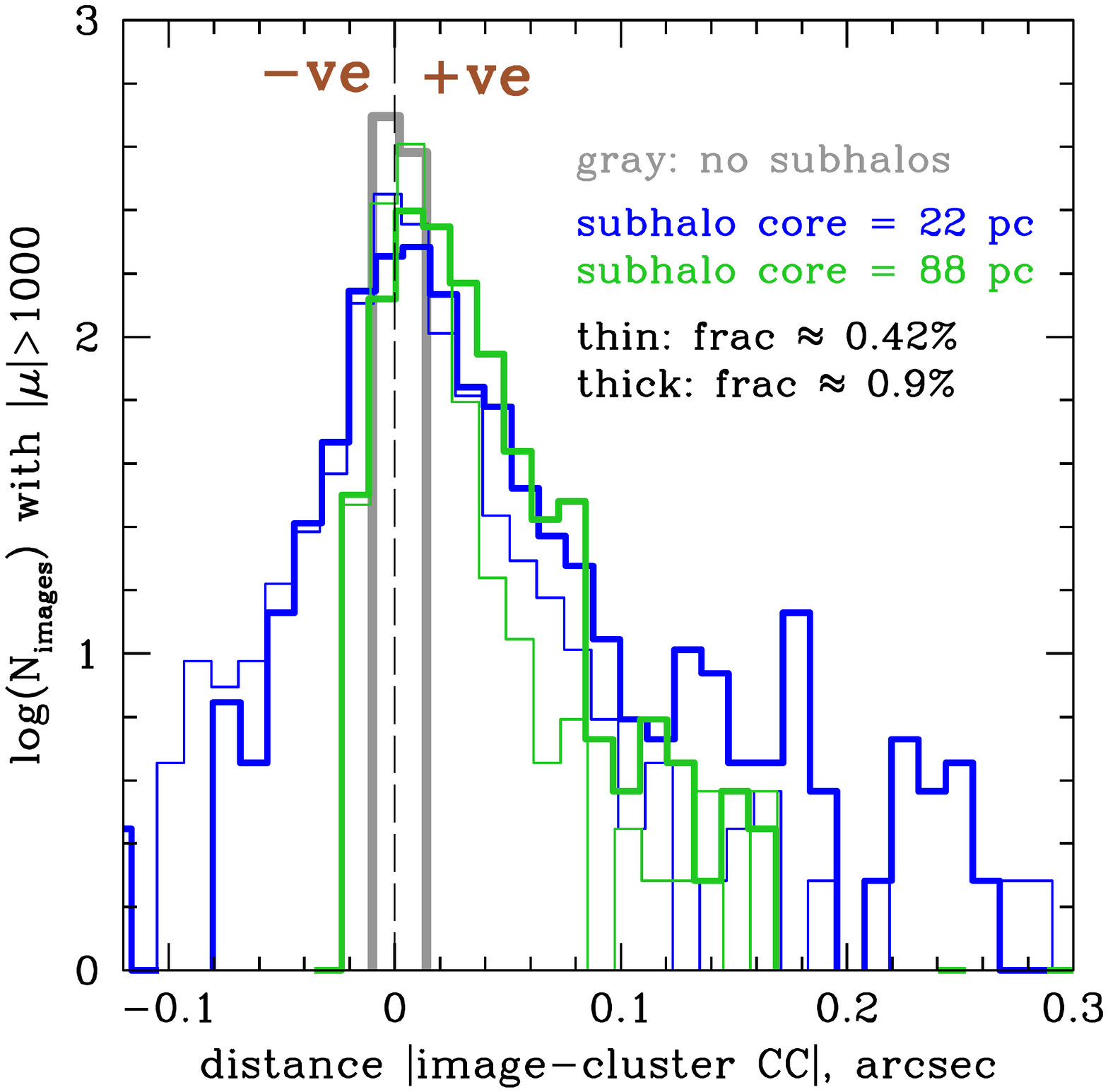}    
    \vskip-2cm
    \caption{The distribution of highly magnified images, $|\mu|>1000$, on the $-$ve side (negative side of the $x$ axis), and +ve side (positive side of the $x$ axis) of the cluster CC. Two different subhalo core radii (blue, 22\,pc; green, 88\,pc), and two different subhalo mass fractions (thin, 0.42\%; thick, 0.9\%) are shown. Sixteen realizations of each case were added to generate these histograms.}
    \label{fig:countnearCC}
\end{figure}

Lens-plane image distributions like Figure~\ref{fig:imageplane} can be summarized by plotting the distribution of images with $|\mu|>1000$ as a function of their distance from the cluster CC, as we do in Figure~\ref{fig:countnearCC}. The distance on the horizontal axis is from the cluster CC; negative (positive) distances are on the $-$ve (+ve) side of cluster.  The blue (green) histograms show cases with subhalo core radii $r_c=22$pc ($88$\,pc). The thin (thick) lines show cases with subhalo mass fraction of $0.42\%$ ($0.9\%$), while the gray histogram represents the case with no subhalos, and is a much narrower distribution than either the blue or green histograms.  The +ve cluster side has more highly magnified images, with some lying further away from the CC: both blue and green histograms extend to further positive than negative distances. The asymmetry is especially pronounced in the green histogram.

This asymmetry, as well as the effect of subhalo core size on the properties of image (blue vs. green histograms in Figure~\ref{fig:countnearCC}), will be discussed in Section~\ref{sec:reasons}. 

\subsection{Magnification Distribution of Sources and Images}\label{sec:sourceimageLF}

\begin{figure}
\includegraphics[trim={0cm 5cm 0cm 4cm},clip,width=0.49\textwidth]{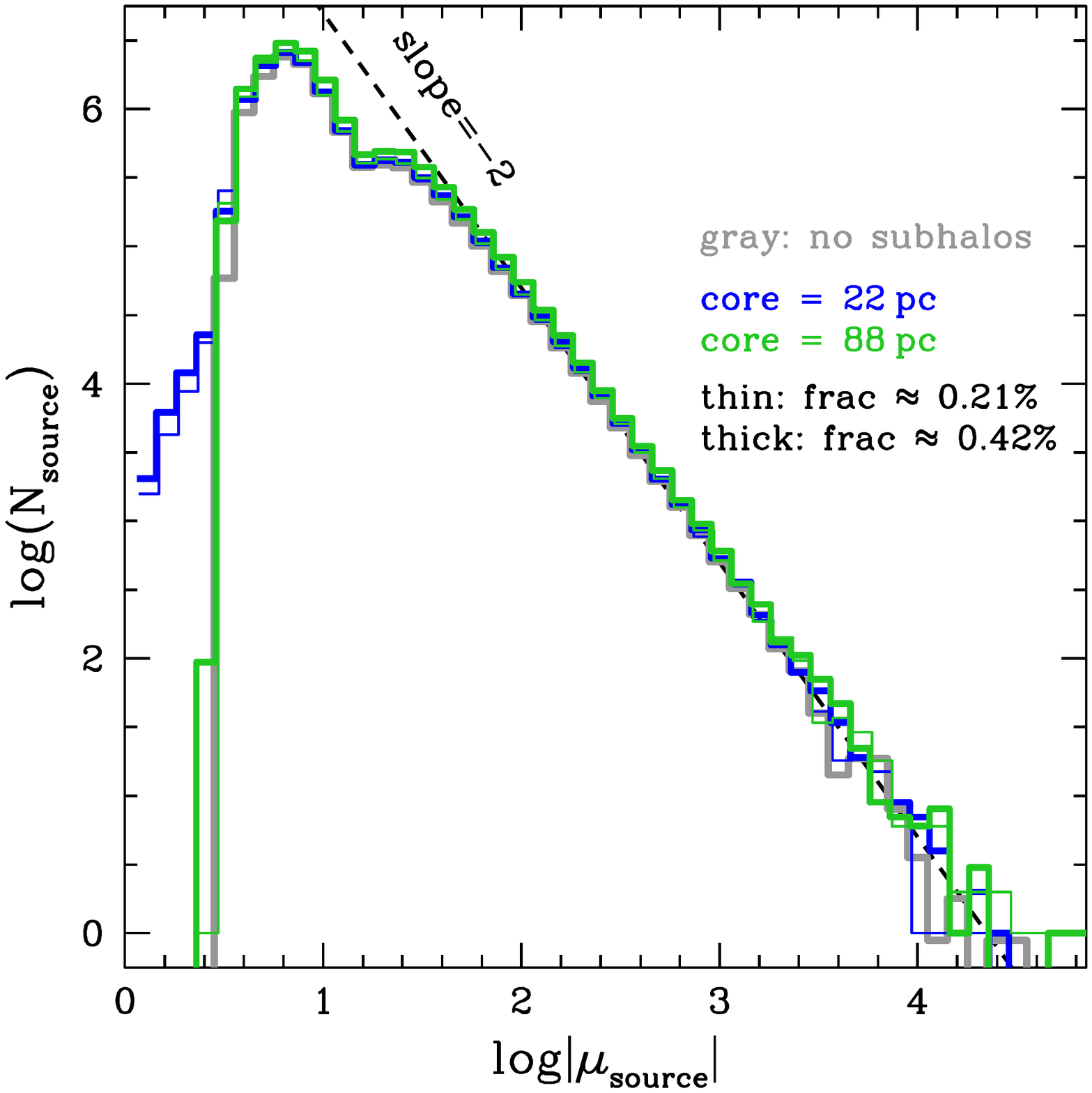}  
\caption{Source magnification distribution. Source magnification is the sum of unsigned image magnifications that end up in our $2\,\square''$ lens-plane modeling window. The extra saddles created by subhalos are mostly responsible for the low-magnification side of the distribution. The two green (blue) histograms represent subhalos with core radii $r_c=88\,$pc ($r_c=22\,$pc). The thin (thick) lines show a subhalo mass fraction of $0.21\%$ (0.42\%). The gray line (mostly hidden behind other lines) depicts the case with no subhalos.\\}
\label{fig:sourceLF}
\end{figure}

\begin{figure*}
\includegraphics[trim={0cm 5cm 0cm 4cm},clip,width=0.49\textwidth]{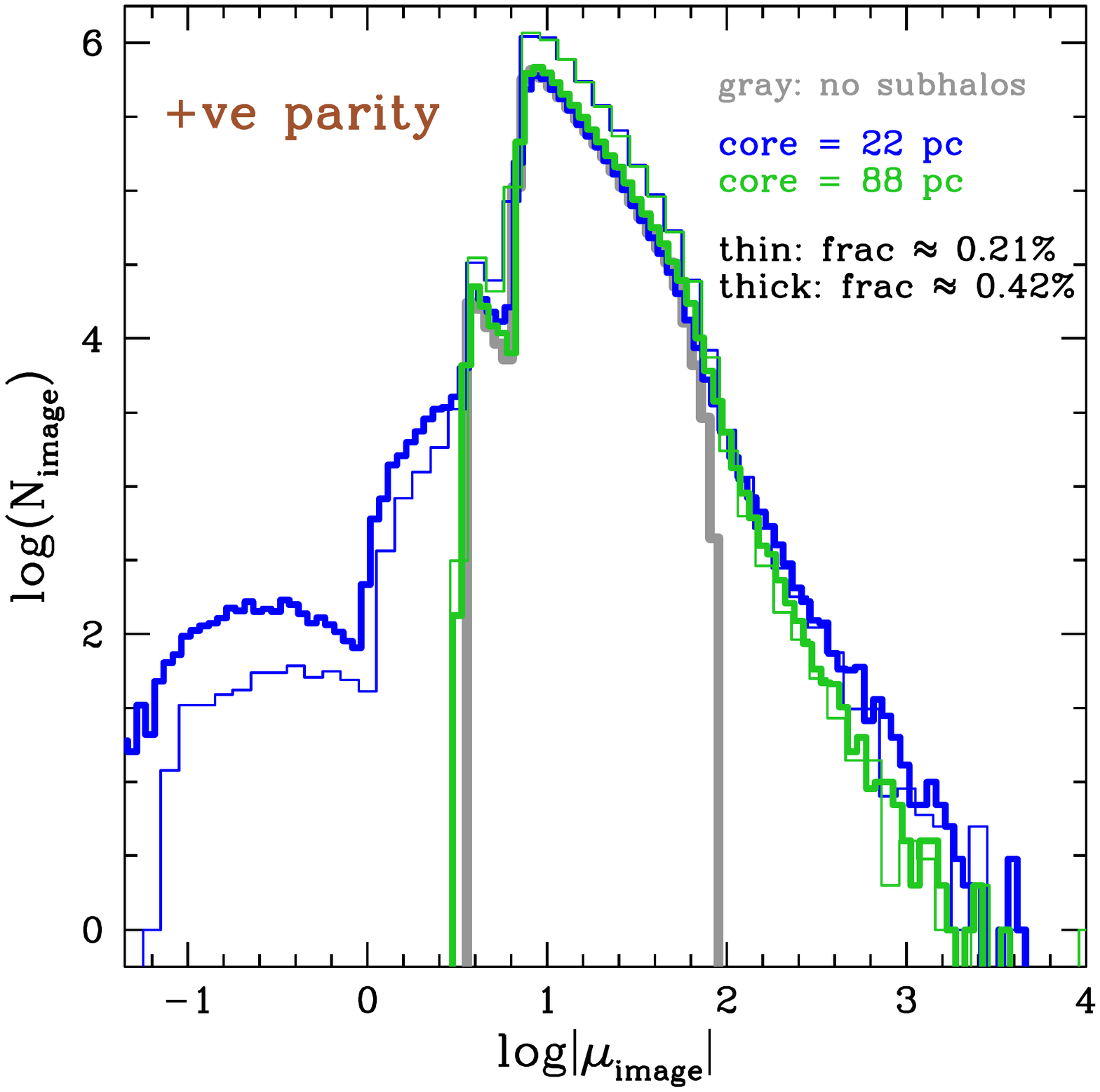}  
\includegraphics[trim={0cm 5cm 0cm 4cm},clip,width=0.49\textwidth]{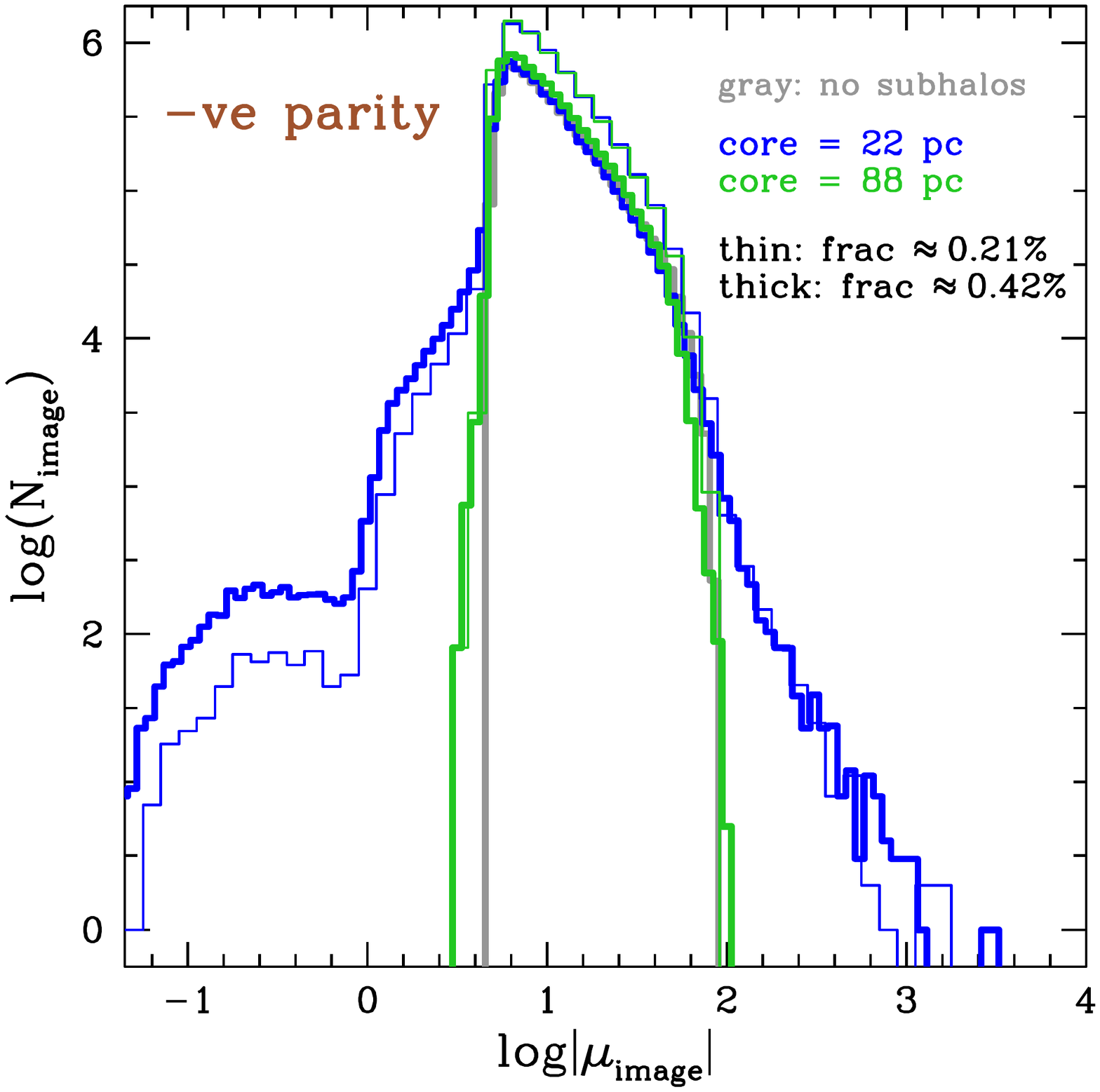}   
\caption{Distribution of unsigned image magnifications in the image plane. The color and line types are indicated in the plot. To avoid the high-magnification images associated with the cluster CC itself, we left out a band of width $\pm 0.1''$ around the cluster CC. {\it Left panel:} Positive parity cluster side. {\it Right panel:} Negative parity cluster side. Subhalos with larger core radii (green histograms) do not produce any very bright images, so the gray (no subhalos) and green histograms nearly overlap.\\}
\label{fig:imageLF}
\end{figure*}

Because of the universal behavior of magnification near caustics, our results should agree with the predictions that the area in the source plane, $\sigma_{\rm sp}$, where sources are magnified by $>|\mu|$ should scale as $\sigma_{\rm sp}\propto |\mu|^{-2}$. The differential distribution of source magnifications, $N_s(|\mu|)d|\mu|$, in the lens plane should follow the same scaling, because
$\sigma_{\rm sp}\propto |\mu|\,\frac{1}{|\mu|}\,N_s(|\mu|)d|\mu|$, where the factor of $|\mu|$ on the right-hand side comes from the fact that we are translating cumulative to differential counts, and the factor of $1/|\mu|$ accounts for the area in the source vs. lens planes.

Figure~\ref{fig:sourceLF} shows the distribution of source magnifications, summed over all the images in our $2\,\square''$ modeling window. The gray histogram represents the case with no subhalos. The blue (green) histograms are for cases with subhalos of $r_c=22\,$pc ($r_c=88\,$pc), and two values of subhalo mass fractions are shown. The high-magnification tail of the distributions, above $|\mu|\approx 100$, is the same for all cases, regardless of subhalo properties. This is the prediction described in the previous paragraph. The faint end of the distribution contains saddle images due to subhalos.

The magnification distribution of images is plotted in Figure~\ref{fig:imageLF}, with the left and right panels showing the +ve and $-$ve cluster sides, respectively. To isolate the effect of subhalos, the images in the immediate vicinity of the cluster CC, $\pm 0.1''$ away from CC,  were excluded.  Because at that distance the typical magnification is $|\mu|\approx 100$,  there is a slight drop in the image number brighter than $|\mu|\approx 100$.  If images closer to the cluster CC were included, their great number would drown out the difference between the +ve and $-$ve sides of the cluster that exists slightly farther away from CC.

The left and right panels of Figure~\ref{fig:imageLF} show images on the {+ve} and {$-$ve} sides of the cluster. Comparing these, it is again apparent that bright images, $|\mu|>100$, are found predominantly on the +ve side of the cluster. 

The most striking feature of Figure~\ref{fig:imageLF} is the difference in the number of highly magnified images generated by subhalos with large core radii (88\,pc) on the +ve vs. {$-$ve} sides of the cluster: compare the green distributions in the left and right panels. While small core radii (22\,pc) subhalos generate fewer images on the $-$ve side compared to the +ve side by a factor of $\sim 2$--3, the large core radii subhalos generate almost no images above some maximum magnification on the $-$ve side. In this figure, that maximum magnification is $|\mu|=100$ (because we removed a band of $\pm 0.1''$ around the cluster CC). In fact, on the $-$ve side, the number of bright images in the case with large-core subhalos and with no subhalos is about the same (gray and green histograms in the right panel). The reason for the absence of subhalo magnified images on the $-$ve side of the cluster in the case of large core radii is discussed in Section~\ref{sec:reasons3}. 

The image magnification distributions have two distinct bumps on the faint side of the peak. The images in the fainter bump are all positive parity images, all located inside the central, smaller critical curve near the centers of many subhalos, on both the +ve and $-$ve cluster sides. The somewhat less faint bump consists predominantly of negative parity images inside the larger, outer CC of subhalos, also on both sides of the cluster CC.  These faint images are not the focus of our work.

From the image distribution we see that about $10^{-5}$ of all images are highly magnified, with $|\mu|>1000$. The fraction is somewhat higher for subhalos with smaller core radii (22\,pc vs. 88\,pc), and somewhat higher for subhalo mass fraction of $\sim 0.5\%$ compared to $\sim 0.25\%$. But these differences are not more than a factor of 2.

\subsection{Number of Highly Magnified Images as a Function of Subhalos' CC Length}\label{sec:reasons2}

Highly magnified images appear almost exclusively near critical curves. Therefore, the number of images highly magnified by subhalos should depend on the total length of the subhalos' CC. This is suggested by Figure~\ref{fig:imageplane}, where the highly magnified images tend to be located on the subhalo CC (gray loops). Here we quantify this visual impression, and in Section~\ref{sec:reasons} we discuss the underlying reasons for why subhalo CC are longer on +ve cluster side.

\begin{figure}
    \vspace{-2cm}
    \centering
    \includegraphics[width=0.49\textwidth]{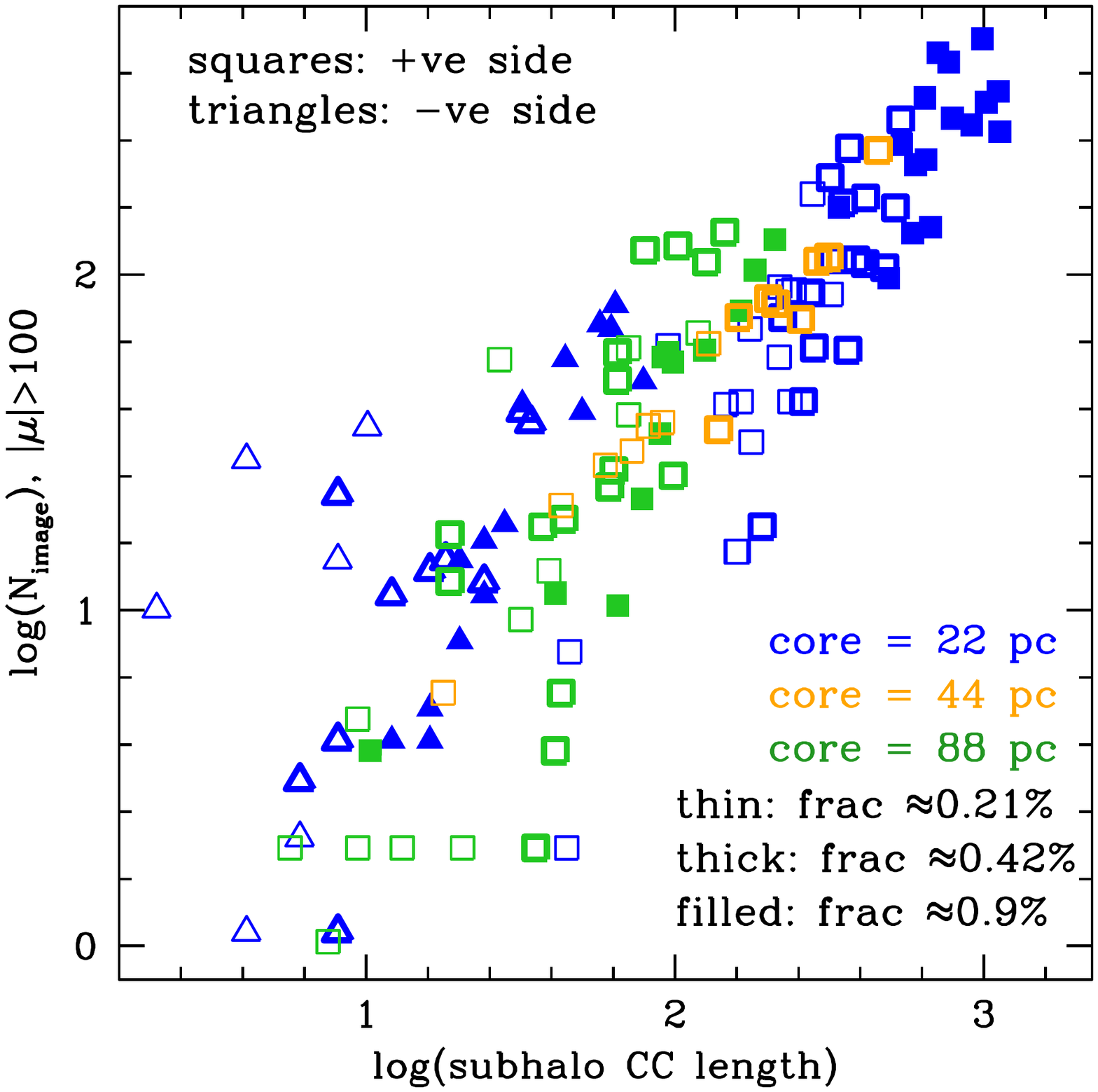}
    \vskip-2.2cm  
    \caption{Log of the number of magnified images with $|\mu|>100$ vs. log of the subhalo CC length (in code units). To isolate the effect of subhalos, the band of width $\pm 0.2''$ around the smooth cluster CC was excluded. Each point represents one realization, and there are between 8 and 16 realizations for most of the cases. If fewer points of a given type appear on the figure, that means either the subhalo length or the number of images with $|\mu|>100$ was zero, so these runs were omitted from the plot. We use three values for core radii (22, 44, and 88\,pc), and three values for subhalo mass fractions ($\sim 0.21\%$, $\sim 0.42\%$, and $\sim 0.9\%$), as indicated in the legend. The points roughly follow a diagonal.}
    \label{fig:subhaloCC}
\end{figure}

Figure~\ref{fig:subhaloCC} plots the log of the total length of all subhalos' CC (in code units) on the $x$ axis vs. the total number of images with $|\mu|>100$ on the $y$ axis. This is done separately for +ve (square symbols) and $-$ve (triangles) sides of the cluster. To capture the effect of just the subhalo CC, without including images magnified with the help of the cluster CC, the band of width $\pm 0.2''$ immediately surrounding the smooth cluster CC was excluded.  \llrw{This exclusion band is made wider here than the $0.1''$ band we used earlier, in order to concentrate specifically on the subhalo CCs that form closed loops, and do not merge with the cluster CC.}
The lower limit of image magnification, $|\mu|>100$, was chosen as a compromise: for lower $|\mu|$, images are not sufficiently magnified, while for higher $|\mu|$, the number of images drops rapidly (see Figure~\ref{fig:imageLF}), resulting in poor statistics. 

Figure~\ref{fig:subhaloCC} shows that the length of subhalos' CC is shorter on the $-$ve cluster side: for any given parameter set, triangles are mostly to the left of squares. The number of magnified images tracks the CC length well, with approximately the same linear relation on the +ve and $-$ve sides. (We constructed a similar plot using $|\mu|>300$ images, which had typically, about 10 times fewer images, but the linear trend was still evident.)

The length of the subhalo CC, \new{which is closely related to the size of the caustic in the source plane,} is the main (though probably not the only) factor that determines the number of highly magnified images. The factors that determine the length of subhalos' CC are discussed in Section~\ref{sec:reasons}. Because the CC length is larger on the +ve side, very bright images are found preferentially on the +ve side of the cluster.

\section{Understanding the Abundance and Distribution of Images as a Function of Subhalo Properties}\label{sec:reasons}

\begin{figure*}
    \centering
    \vspace{-1.5cm}
    \includegraphics[width=0.495\textwidth]{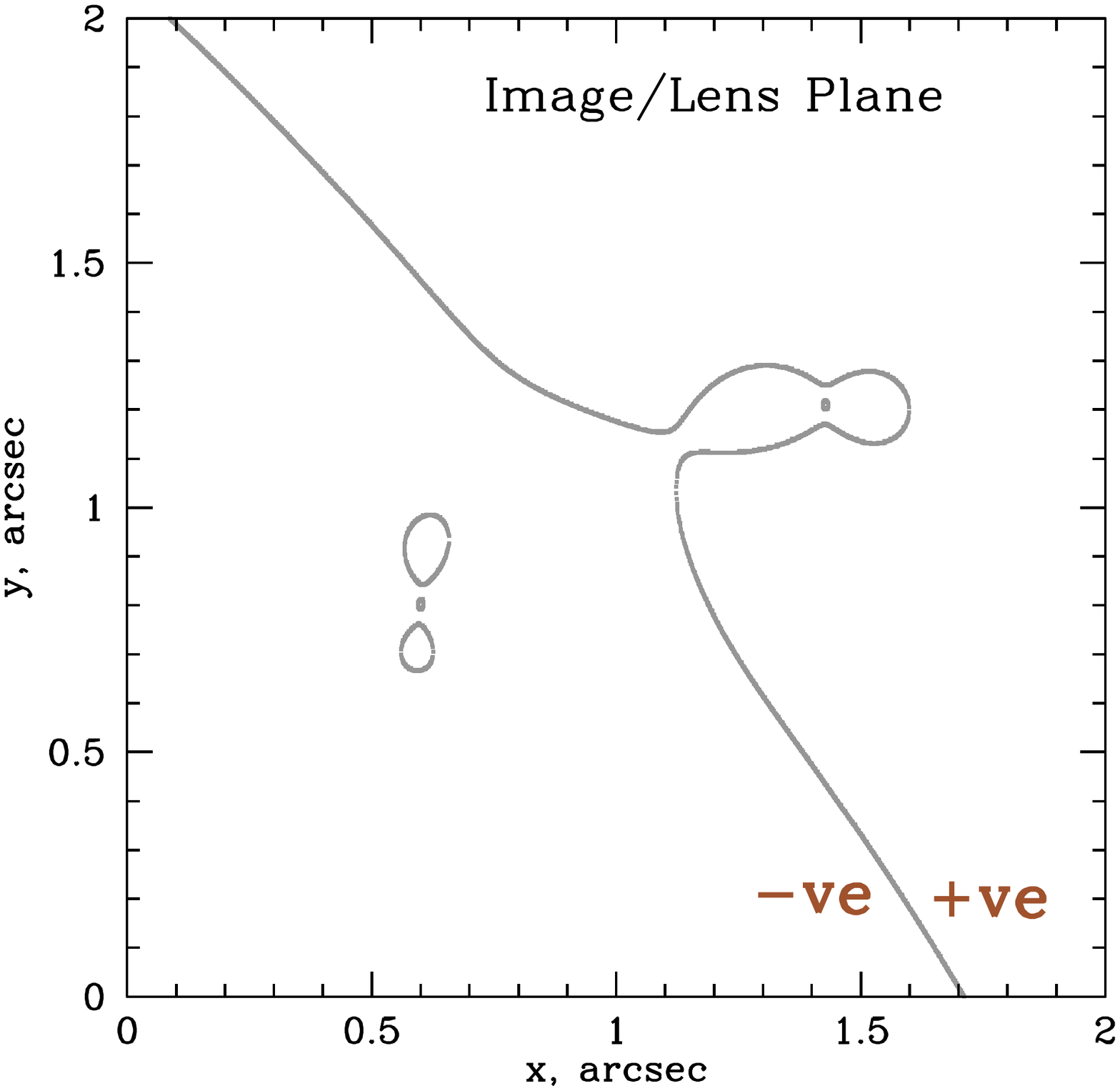}
    \includegraphics[width=0.495\textwidth]{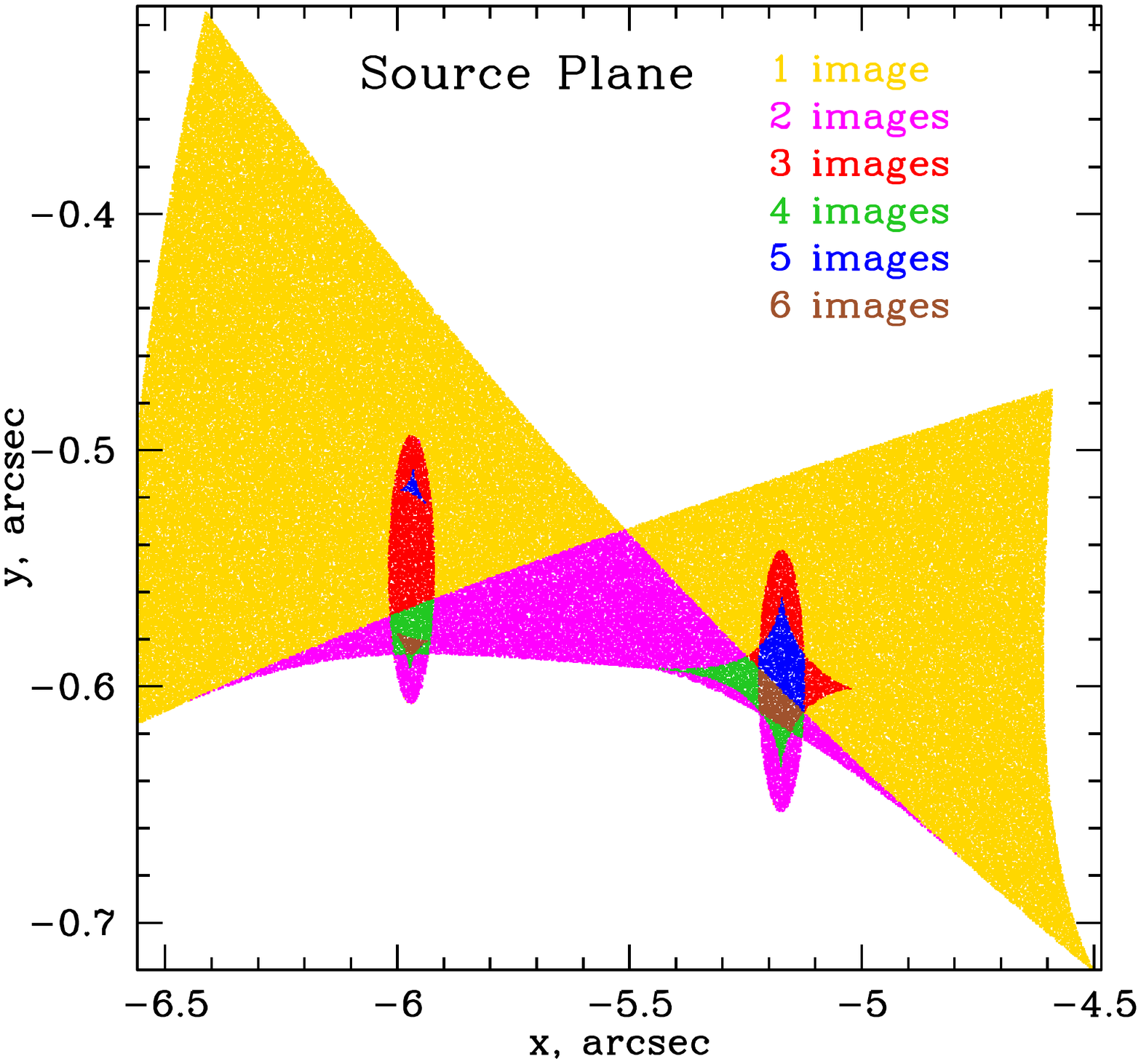}
    \vskip-2.3cm
    \caption{An illustration of subhalos' CC on the +ve and $-$ve sides of the cluster. (These runs were not used for the actual calculations.) {\it Left panel:} Image, or lens plane, with two identical subhalos equidistant from the cluster CC. As discussed in Section~\ref{sec:reasons}, lensing rotates the orientation of the subhalos by $\sim \pi/2$ on the opposite sides of the cluster CC, and size is smaller on the $-$ve side owing to the density gradient of the main cluster. {\it Right panel:} The portion of the source plane from which the images in the left panel would originate. The six different colors show regions giving rise to six different image multiplicities. Note that only those images that would appear in the image-plane square of the left panel are counted. This figure is for illustration purposes only; these subhalos are more massive than those used in the main calculations. Subhalos used in calculations rarely generate five or six image configurations.\\}
    \label{fig:dir51}
\end{figure*}

\subsection{General Considerations}\label{sec:reasons1}

Since we are considering image properties near the CC, it is important to discuss the asymmetry between the +ve and $-$ve cluster sides, which affects image properties. This is addressed analytically by \cite{ven17}, \cite{ogu18}, and \cite{die18} for microlensing, but these papers do not consider the situation where the macro density due to the cluster is different on the two sides of the cluster CC, and the effect of the subhalo core size. Here, we investigate these properties. \cite{dai20} and \cite{kau19} do simulations of cored subhalos, but do not explore the effect of these properties.

In the following subsections we discuss how image properties are affected by the cluster density gradient, the location of subhalos on the +ve and $-$ve sides of the cluster, and the core sizes of subhalos. The combined effect of these factors determines the length of subhalos' CC, and hence the abundance of highly magnified images, as presented in Section~\ref{sec:reasons2}. 

We consider a simple example: two identical subhalos placed at equal distances on the opposite sides of the cluster CC; the resulting critical curves are shown in Figure~\ref{fig:dir51}. Though the subhalos are identical, the critical curves associated with them are different in two respects: on the $-$ve side, the subhalo CC is smaller and forms an upright figure ``8''; on the +ve side, the CC is larger and forms an ``$\infty$" sign. In what follows we will describe how these and other properties arise.

Let the macro cluster potential be a singular isothermal sphere (SIS). Whether the cluster has a flat density core does not matter, since the cluster CC is well outside it, where the density profile is falling as a power law. Its lensing potential is 
\begin{equation}
   \Psi=(x^2+y^2)^{1/2}=r\label{eqn:psi}\, ,
\end{equation}
where we ignore the normalization as it is not important for us. The corresponding dimensionless projected density and shear components are given by
\begin{align}
  \kappa&=\frac{1}{2r}\nonumber \, ,\\
  \gamma_1&=\frac{1}{2}\frac{(y^2-x^2)}{r^3}= -\frac{\cos 2\phi}{2r}\nonumber \, ,\\
  \gamma_2&=\frac{1}{2}\frac{(xy)}{r^3}= -\frac{\sin 2\phi}{2r}\, .
  \label{eqn:params}
  \end{align}
The total shear amplitude due to SIS is given by $\gamma=\sqrt{(\gamma_1^2+\gamma_2^2)}$. SIS has $\kappa=\gamma$ at all radii, and both the projected density and shear fall as $1/r$. The two eigenvalues of the magnification matrix are $(1-\kappa-\gamma)=(1-2\kappa)$, and $(1-\kappa+\gamma)=1$; hence, the magnification of an image at a distance $r$ from the center is $(1-2\kappa)^{-1}=r/[r-1]$. The critical curve is at radius $r=1$, and the corresponding density is $\kappa_{\rm CC}=0.5$. \llrw{The orthogonally-oriented figure ``8'' shapes of the iso-$\gamma_1$ and iso-$\gamma_2$ contours are discussed in Section~\ref{appC}.} 

The orientation of our Cartesian coordinate system is shown in Figure~\ref{fig:images00}. The cluster CC stretches the image almost exactly vertically, along the $y$ axis. This makes the analysis here easier, but the conclusions obtained here hold in general.

\subsection{The Effect of Cluster Density Profile Gradient}\label{sec:gradient}

The density profile of the cluster plays an important role in the size of the subhalo CC. \new{Let us assume that the cluster density profile is a power law, $\kappa(r)=\kappa_{\rm CC}\,r^{-\alpha}$, where $r$ is the distance away from the cluster center in units of its CC distance, so at CC, $r=1$, and the projected density at that location is $\kappa_{\rm CC}$. The density at a distance $\Delta r$ away from the cluster CC (i.e., at a distance $r=1+\Delta r$ from the cluster center) can be obtained using Taylor expansion:
\begin{align}
    \frac{\kappa(r)}{\kappa_{\rm CC}}&=r^{-\alpha}\nonumber\\
    &= r^{-\alpha}|_{\Delta r=0}-\alpha\, r^{-1-\alpha}|_{\Delta r=0}\,\Delta r\nonumber\\
    &+\frac{1}{2!}\alpha(1+\alpha)r^{-2-\alpha}|_{\Delta r=0}\,[\Delta r]^2+...\nonumber\\
    &\approx 1-\alpha\,\Delta r + \frac{1}{2}\alpha(1+\alpha)\,[\Delta r]^2\, ,
    \label{eq:taylor0}
\end{align}
where $\Delta r$ can be positive or negative. For a SIS which has $\alpha=1$, this becomes 
\begin{equation}
    \kappa(r)\approx\kappa_{\rm CC}\times(1-\Delta r+ [\Delta r]^2)\, .
    \label{eq:taylor1}
\end{equation} 
At very small distances from the CC, the $[\Delta r]^2$ term is negligible, and $\kappa$ can be represented by a constant-density slope. But for somewhat larger displacements from the CC, those relevant for subhalos, the $[\Delta r]^2$ term is important, and it makes the slope nonlinear. }

On the +ve side, $\Delta r>0$, 
\begin{equation}
   \Delta\kappa_+=\kappa(1+\Delta r)-\kappa_{\rm CC}=\kappa_{\rm CC}\times(-\Delta r + [\Delta r]^2)\, ,
\end{equation}
while on the $-$ve side, $\Delta r<0$,
\begin{equation}
   \Delta\kappa_-=\kappa(1-\Delta r)-\kappa_{\rm CC}=\kappa_{\rm CC}\times(|\Delta r| + [\Delta r]^2)\, ,
\end{equation}
implying that $|\Delta\kappa_-|>|\Delta\kappa_+|$ for the same value of $|\Delta r|$ on both sides of the cluster CC.
In other words, $\kappa$ rises faster away from the CC on the $-$ve side, compared to its fall on the +ve side, consistent with the power law in the linear-linear representation.

With this we can rewrite Eq.~\ref{eq:tanCCapprox} of Section~\ref{appC} for the +ve side as 
\begin{equation}
0=1-2\kappa_{\rm CC}(1-\Delta r+[\Delta r]^2)-|\gamma_{1s}|=\Delta r-[\Delta r]^2-|\gamma_{1s}|\, ,\nonumber
\end{equation}
where the last expression assumed that $\kappa_{\rm CC}=0.5$. (We use $-|\gamma_{1s}|$ to explicitly state that that term is negative.) 
Or, equivalently, 
\begin{equation}
0=-\Delta r+[\Delta r]^2+|\gamma_{1s}|\, .
\label{eq:slopepos}
\end{equation}

For the $-$ve side, the corresponding equation is 
\begin{equation}
0=1-2\kappa_{cc}(1+|\Delta r|+[\Delta r]^2+\gamma_{1s})=-\Delta r-|\Delta r|^2+\gamma_{1s}\, .
\label{eq:slopeneg}
\end{equation}
In Eq.~\ref{eq:slopeneg}, the $\gamma_{1s}$ term has to compensate for two negative terms to satisfy the equation, while in Eq.~\ref{eq:slopepos}, $|\gamma_{1s}|$ has to compensate for a smaller value. Therefore, $\gamma_{1s,\rm Eq.\ref{eq:slopeneg}}>|\gamma_{1s,\rm Eq.\ref{eq:slopepos}}|$. Larger values of shear are closer to the center of the subhalo, implying that on the $-$ve cluster side, subhalos' CC will be smaller in size. Note that this size asymmetry is due to the presence of the $[\Delta r]^2$ term, which has different signs in the two equations. In a cluster with a linear density profile instead of a power law, this asymmetry in subhalo CC size would not exist.

The asymmetry is seen in the left panel of Figure~\ref{fig:dir51}. It is also present in the subhalos' signature on the +ve vs. $-$ve parity sides of the cluster in Figure~S4 (p. 48) of \cite{men20}, which shows caustics on both sides of the cluster CC.

It follows that steeper power-law cluster density profiles will result in larger disparity between subhalo CC sizes on the two sides of cluster CC. On the other hand, if the cluster has a linear density slope, or the density is approximately the same on both sides, then subhalo CC sizes will be roughly the same, which is seen in Figure 4 (right panel) of \cite{die18}.

To quantify the effect of shallower power-law cluster density profiles near the locations of the CC, we also do our simulations for a $\sim {r_{2D}}^{-0.9}$ profile, and present these results in Appendix~\ref{appB}.

\subsection{The Effect of Subhalo Core Size}\label{sec:reasons3}

The case of a large subhalo core on the +ve vs. $-$ve sides can be addressed with a simple analytical argument based on the discussion earlier in this section. \new{In Section~\ref{sec:gradient}, we considered a subhalo density profile with a small concentrated core, and saw that there exist solutions to Eqs.~\ref{eq:slopepos} and \ref{eq:slopeneg} for subhalo CC on either side of the cluster CC. In the case of a subhalo with a large flat density core, $\gamma\ll\kappa$, and there may not be a solution to Eq.~\ref{eq:slopeneg}, implying that subhalos may not even form their own CCs on the {$-$ve} side. On the {+ve} side, on the other hand, it would be easier to have solutions to Eq.~\ref{eq:slopepos}, for the same subhalo properties.}

Absence of a subhalo CC for large core radii, and on the $-$ve cluster side, means that there will not be any highly magnified images there. That is why there are no orange or green triangles ($r_c=44\,$pc and $88\,$pc; $-$ve side) in Figure~\ref{fig:subhaloCC}. 

\begin{figure}
    \centering
    \vspace{-2cm}
    \includegraphics[width=0.495\textwidth]{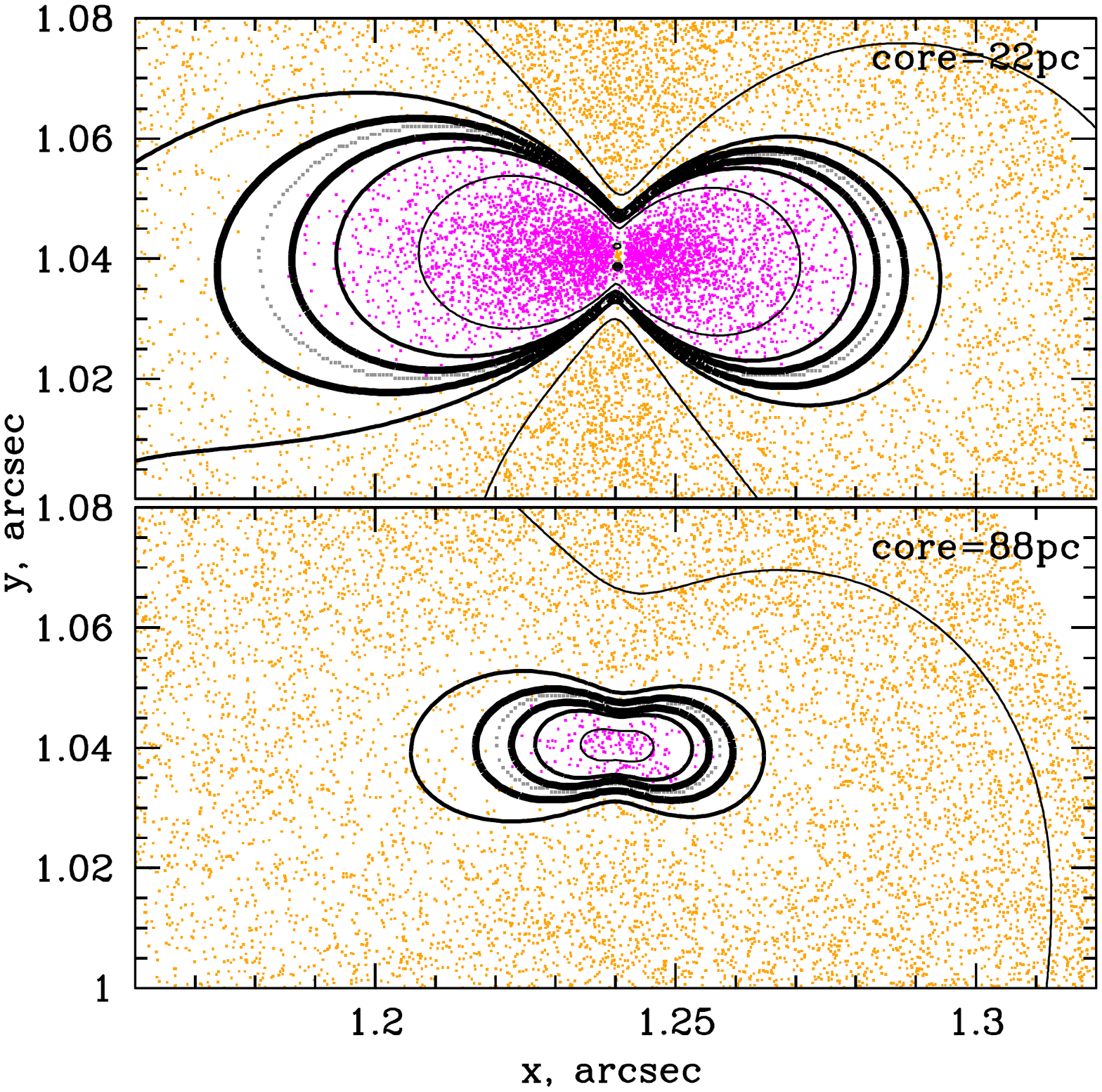}
    \vskip-2.2cm
    \caption{Zoom-in regions around two individual isolated subhalos, in two separate controlled runs. Both are on the +ve side of cluster CC. Both subhalos have the same total mass, but different core radii. {\it Top:} Subhalo core radius = 22\,pc; {\it Bottom:} Subhalo core radius = 88\,pc. The critical curves are the gray dotted lines; images of positive and negative parity are shown as orange and magenta points, respectively. The density of source in the source plane is the same in both cases. The three sets of black curves show unsigned magnifications of 300 (thick), 100, and 30 (thin). The dilution of images in regions of high magnification is apparent in both cases. \\}
    \label{fig:DEMOimageplane}
\end{figure}

Even on the +ve side, the subhalo core size will affect the extent of its CC.  To demonstrate that we perform a numerical experiment. We generate two new runs, each with one subhalo of the same mass, but different core radii, on the +ve cluster side, and at the same distance from the cluster CC. The same number of background point sources is used in both cases. Figure~\ref{fig:DEMOimageplane} shows the results: the subhalos in the top and bottom panels have core radii of $22\,$pc, and $88\,$pc, respectively. Curves of equal magnification are also shown, as well as the images of +ve (orange) and $-$ve (magenta) parity. The larger core radius results in the subhalo's mass getting more spread out, and its CC and high magnification regions smaller, leading to fewer highly magnified images compared to a smaller core radius case. This is why in Figure~\ref{fig:subhaloCC} the blue squares ($r_c=22\,$pc; +ve side) are to the upper right of the orange squares ($r_c=44\,$pc; +ve side), which are in turn to the upper right of the green squares ($r_c=88\,$pc; +ve side). 

\subsection{The Effect of the Shape of the Arrival Time Surface}\label{sec:arriv}

\begin{figure*}
    \centering
    \includegraphics[trim={0cm 5cm 0cm 4cm},clip,width=0.495\textwidth]{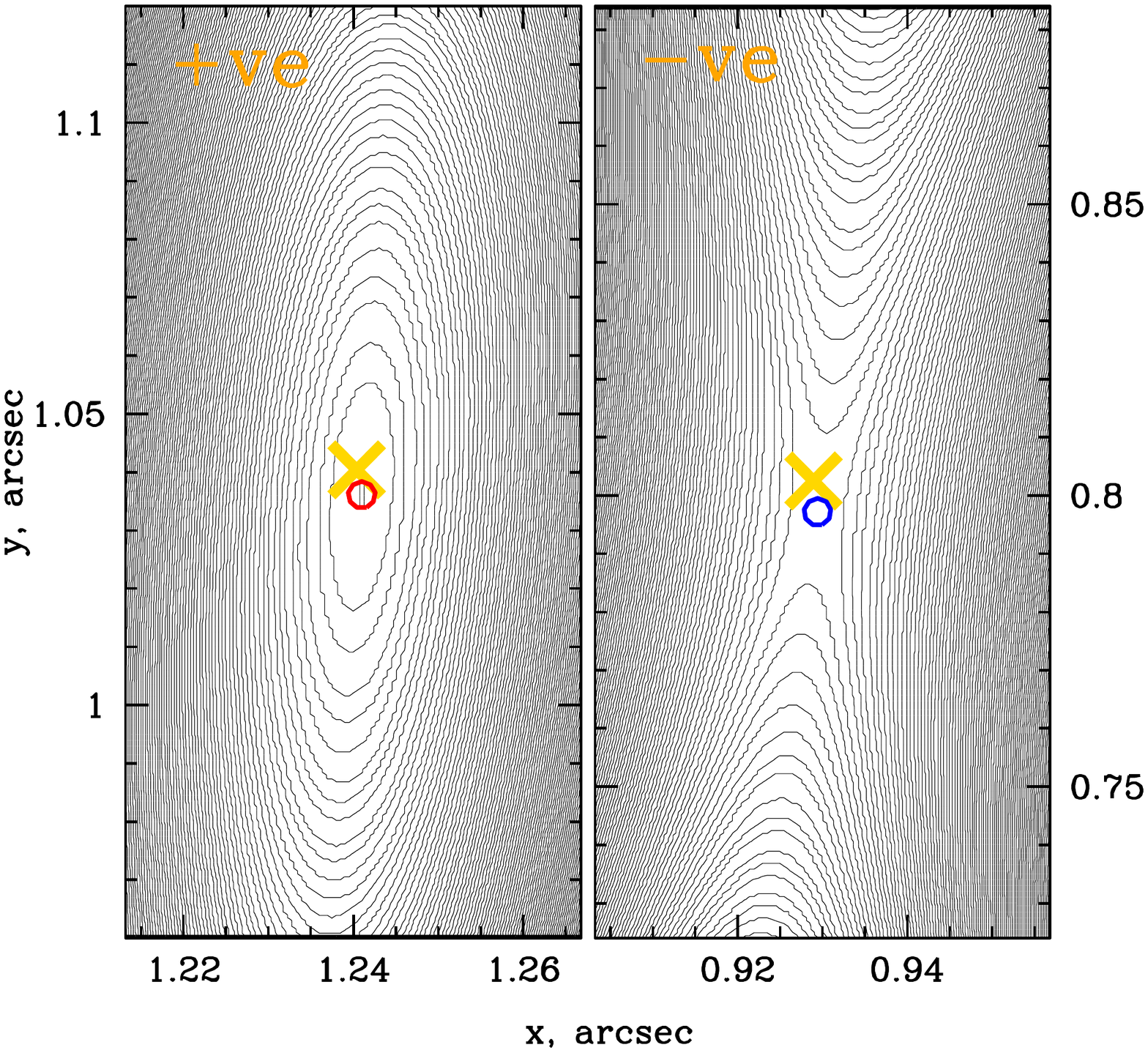} 
    \includegraphics[trim={0cm 5cm 0cm 4cm},clip,width=0.495\textwidth]{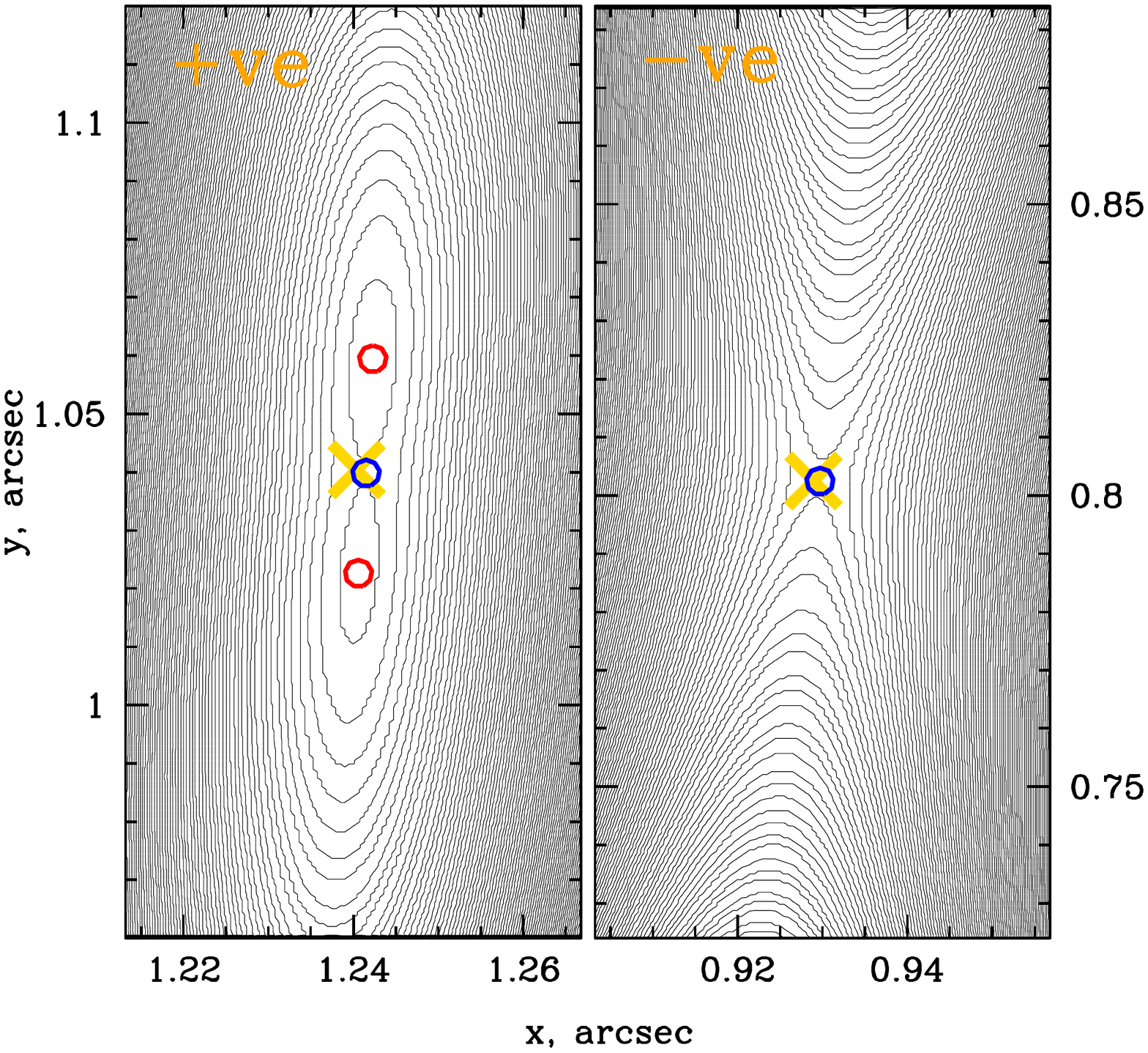}
    \caption{Arrival time surface around the locations of individual subhalos, indicated by yellow crosses. Positive and negative parity images are red and blue circles, respectively. The left and right sides within the left and right panels show the +ve and $-$ve sides of the cluster, as marked on the plots. {\it Left panels:} Subhalo mass set = 0, to show the location of images for a subhaloless cluster. The magnifications of the two images are 29.86 (red circle) and $-27.87$ (blue circle). {\it Right panels:} Each subhalo has mass of $10^7\, M_\odot$. The magnification of the $-$ve parity image on the $-$ve side (blue circle) is $-7.48$, while the total unsigned magnification of the three images of the lemniscate on the +ve side is 93.01. To facilitate comparison, contour levels in all four panels are the same values, \new{and are spaced by about an hour in the observer frame.}\\}
    \label{fig:smlpotPATCH}
\end{figure*}

In the left panel of Figure~\ref{fig:smlpotPATCH} we show
a smooth substructureless cluster that generates images of a point source, of opposite parity on either side of its CC. The unsigned image magnifications are approximately the same. Black contours are those of the arrival time surface of two sources, whose images are the locations marked with red (+ve parity image), and blue ($-$ve parity image) circles. Since this is a subhaloless cluster, +ve ($-$ve) parity image is on the +ve ($-$ve) side of the cluster.  The unsigned magnifications of these are about equal, as expected, 29.86 and 27.87. 

In the presence of substructures, either subhalos or microlenses that are concentrated enough to form two extra images, the +ve side will add a lemniscate contour in the time-delay surface with a minimum and a saddle, or a lemacon contour with a maximum and a saddle \citep[see Figure 1 of][]{sah11}. The $-$ve side will add a maximum and a saddle \citep[see upper left in Figure 2 of][]{sah11}. Because minima are usually not demagnified below the magnification due to the smooth cluster, while saddles and maxima can and often are, the +ve side of the cluster will generally acquire more magnified images than the $-$ve side. 

In the case of subhalos, which differ from microlenses by having an extended mass distribution and possibly a flat density core, an additional effect is in play, which makes minima on the +ve side even more magnified.

The right panels of Figure~\ref{fig:smlpotPATCH} use the same sources, and the same patches of the lens plane, but now introduce one subhalo at each of the locations marked by yellow crosses. The mass of each is $10^7\, M_\odot$, and the core radius = 88\,pc. On the +ve side of the cluster a subhalo forms a lemniscate configuration with three images, as stated above. Since the contour levels are drawn at the same value for all four panels, their spacing can be used to judge the relative magnification of images. The shallowness of the arrival time surface on the +ve side indicates that the images will appear bright; in fact the total magnification of the three images is 93.01, compared to $\mu=29.86$ on the +ve cluster side image in the left panel of Figure~\ref{fig:smlpotPATCH}. The shallowness of the arrival time can be explained as follows.

On the +ve side of the subhaloless cluster the arrival time surface around an image looks like a bowl, and so hosts minima. Adding a nearby subhalo makes the profile of that bowl shallower (with or without creating extra images), resulting in higher magnification. 

The situation on the $-$ve parity side of the cluster is different: the shape of the arrival time is that of a saddle, and so hosts saddle images. Adding a smooth bump due to a subhalo does not flatten out the arrival time shape, and hence does not increase the image magnification. This is seen in the spacing of the contours on the $-$ve side of the two panels. In fact, the image magnification in the presence of a subhalo (right side of the right panel) is lower ($\mu=-7.48$) than that of the same image when the subhalo is absent ($\mu=-27.87$).

We conclude that on the +ve side of the cluster, subhalos, which have an extended mass distribution, and possibly a flat density core, will generally magnify images more than point-mass microlenses would alone. The exception is when microlenses form more than two extra images on the $-$ve side, some of which will be minima \citep[see the other three panels in Figure 2 of][]{sah11}.

\section{Observable Signatures of Subhalos}

Our goal in this section is to identify tools to detect the presence of dark matter subhalos near the cluster CC, and to estimate their properties. We use subhalos only in the mass range $10^6$--$10^8\,M_\odot$; more massive subhalos will likely produce a stronger signature. We describe two different types of subhalo signatures: their effect on the pixelated flux distribution (Section~\ref{sec:pixel}), and the effect of individual highly magnified images (Section~\ref{sec:indiv}).  In this paper we work with simulations only, but plan to extend the method to observations of clusters, specifically from the ongoing {\it HST} Flashlights project.

\subsection{Pixel-Level Signatures}\label{sec:pixel}

First, we devise a metric suitable for working with flux collected in detector pixels, with an applied point-spread function (PSF), which we model as a Gaussian with a width of $\sigma=0.04''$.  \new{(A similar problem has recently been analyzed by \cite{bay23} for subhalos in galaxy-scale lenses.)} We bin the flux from images into pixels, and then apply PSF convolution of width $0.04''$, comparable to that of optical {\it HST} PSFs.\footnote{In principle, one would first convolve each image with a PSF and then bin the resulting flux distributions, but we reversed the order of these two operations for computational ease. \llrw{We repeated the calculation for a subset of runs, where we first applied PSF convolution and then binned the flux; the results were very close to the ones presented here.}}

\subsubsection{Local Flux Peaks and Holes}

\begin{figure}
\includegraphics[trim={0cm 5cm 0cm 4cm},clip,width=0.49\textwidth]{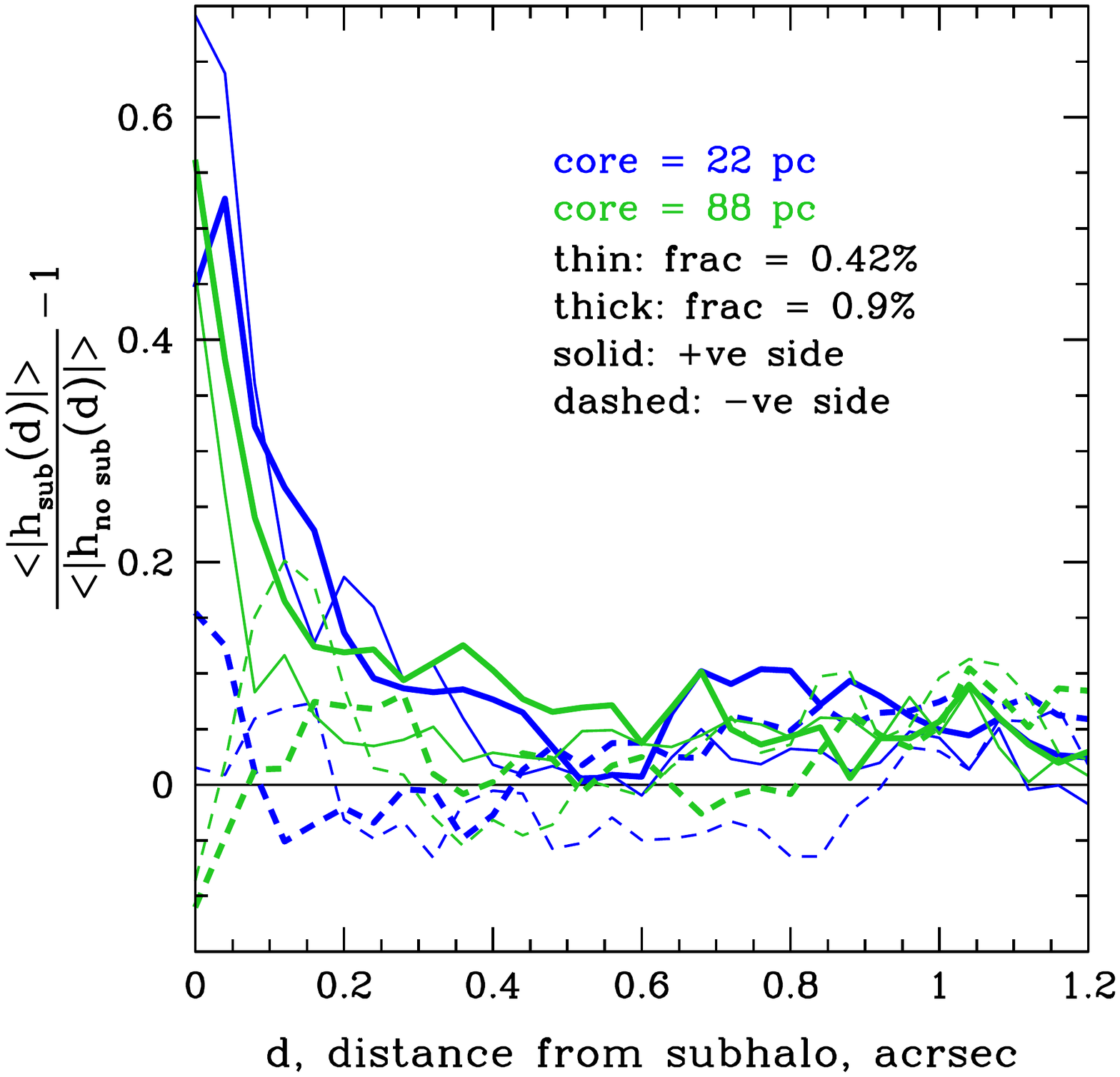}
\caption{Average value of the pixels' local flux peak height or hole depth, $|h|$ (Eq.~\ref{eq:h}) in the presence of subhalos, compared to the case where subhalos are absent, as a function of the distance between it and subhalos (Eqs.~\ref{eq:ave_h} and \ref{eq:correl}).
The average is over all the subhalos outside the $\pm 0.1''$ band around the smooth cluster CC.  Sixteen individual realizations were used for each line. The legend indicates model parameters. There is a correlation between peaks/holes, and subhalos on the +ve cluster side (solid lines) but not on the $-$ve side (dashed lines).}
\label{fig:correl}
\end{figure}

In Figure~\ref{fig:imageplane} we saw that away from the cluster CC, highly magnified images are strongly associated with the subhalos' CC, on the +ve cluster side. The amplitude of association depends on the subhalos' density profile and the cluster density profile. Because lensing magnifies point sources and at the same time dilutes their number density, the lens-plane number density of highly magnified images is decreased by the same factor as the magnification. Therefore, near subhalos on the +ve side, which generate many highly magnified images with a low number density, we also expect to find considerable fluctuations in pixel fluxes between different lens-plane locations.

To quantify that, we first introduce a metric to quantify local fluctuations, and then describe how we measure their distribution in the lens plane. Flux variations (i.e., the heights of local peaks and the depths of local holes) are calculated as  
\begin{equation}
  h=\log_{10}(f_{\rm in}/f_{\rm out}(p))\, ,    
  \label{eq:h}
\end{equation}
where $f_{\rm in}$ is the flux in a $0.04'' \times 0.04''$ pixel, and $f_{\rm out}$ is the average flux in the surrounding pixels, up to $p$ pixels away. For $p=1$ ($p=3$), the number of surrounding pixels is $(2p+1)^2-1$, i.e., 8 (48).  After experimenting with a few values of $p$ we settled on $p=3$. A value of $h=0.02$ corresponds to a flux ratio $(f_{\rm in}/f_{\rm out})$ of about 1.05 in the central pixel compared to the average of its neighbors, or $-0.05$ mag. Peaks have $h>0$, while holes have $h<0$.

We measure the distribution of peaks and holes as follows. At a distance $d$ from every subhalo, we calculate the average value of $\langle|h|\rangle$, weighted by the mass of the subhalo, 
\begin{equation}
   \langle |h_{\rm sub}|\rangle (d)= 
   \frac{\sum_{\rm sub}\sum_{\rm pix} (|h|\,m_{\rm sub})}
        {\sum_{\rm sub}\sum_{\rm pix} (m_{\rm sub})}\, . \label{eq:ave_h}
\end{equation}
The sum is over all pixels and all subhalos, but leaving out subhalos in a band around the CC of $\pm 0.1''$. We then normalize these by the corresponding values of subhaloless clusters to get a correlation-function-like quantity,
\begin{equation}
    w(d)=\frac{\langle |h_{\rm sub}(d)|\rangle}
              {\langle |h_{\rm no\,\, sub}(d)|\rangle}-1\, . \label{eq:correl}
\end{equation}
In subhaloless clusters, instead of subhalo positions we use randomly generated points in the lens plane. In Figure~\ref{fig:correl} we plot Eq.~\ref{eq:correl}. As in the previous plots, solid lines represent the +ve side of the cluster. These show a positive correlation with subhalos, which means that subhalos are surrounded by local flux peaks and holes of larger amplitude than peaks and holes farther away from subhalos. Dashed lines represent the $-$ve side of the cluster; locations of peaks and holes on the $-$ve side of the cluster do not correlate with subhalo locations. 

The fact that subhalos are surrounded by regions of higher fluctuations in pixel flux is reminiscent of the principle used to measure distance to elliptical galaxies using the surface brightness fluctuation method \citep{ton88}. In those studies, galaxies are assumed to have the same surface brightness, and the reason for some galaxies having fewer but brighter stars, instead of more numerous fainter stars per sky area, is the distance to the galaxies. In the case of lensing, it is the magnification of images coupled with the decrease in their number density on the sky. In both cases the underlying principle is the conservation of surface brightness. 

Since the correlations quantify the average number density profile of peaks and holes around stacked subhalos, we do not expect $w(d)$ in Figure~\ref{fig:correl} to be different for cases of low vs. high subhalo mass fraction, but we might expect a difference between the cases of small and large subhalo core radii. A small difference is visible, with the smaller core subhalos generating somewhat stronger correlations.

The main conclusion from Figure~\ref{fig:correl} is that for a subhalo mass fraction of $\sim 0.5$--1\% (and presumably higher), the amplitude of flux variations between pixels in a radius $\lesssim 0.2''$ around subhalos is significantly higher than at larger distances. Furthermore, such fluctuation will be found predominantly on the +ve cluster side. That means flux variations may signal the presence of dark subhalos on the +ve side; on the $-$ve side, subhalos do not generate higher than typical flux variations.

\subsubsection{Detecting Pixels Affected by Subhalos}\label{sec:obs}

The correlation function, $w(d)$ described above, is one possible way to detect subhalos. Another observable test for subhalos would be to take advantage of the fact that in the presence of subhalos, the +ve and $-$ve sides of the cluster have different degrees of flux variations between pixels. When subhalos are absent, the differences between the two sides of the cluster are small.

\begin{figure*}
\includegraphics[trim={0cm 5cm 0cm 4cm},clip,width=0.49\textwidth]{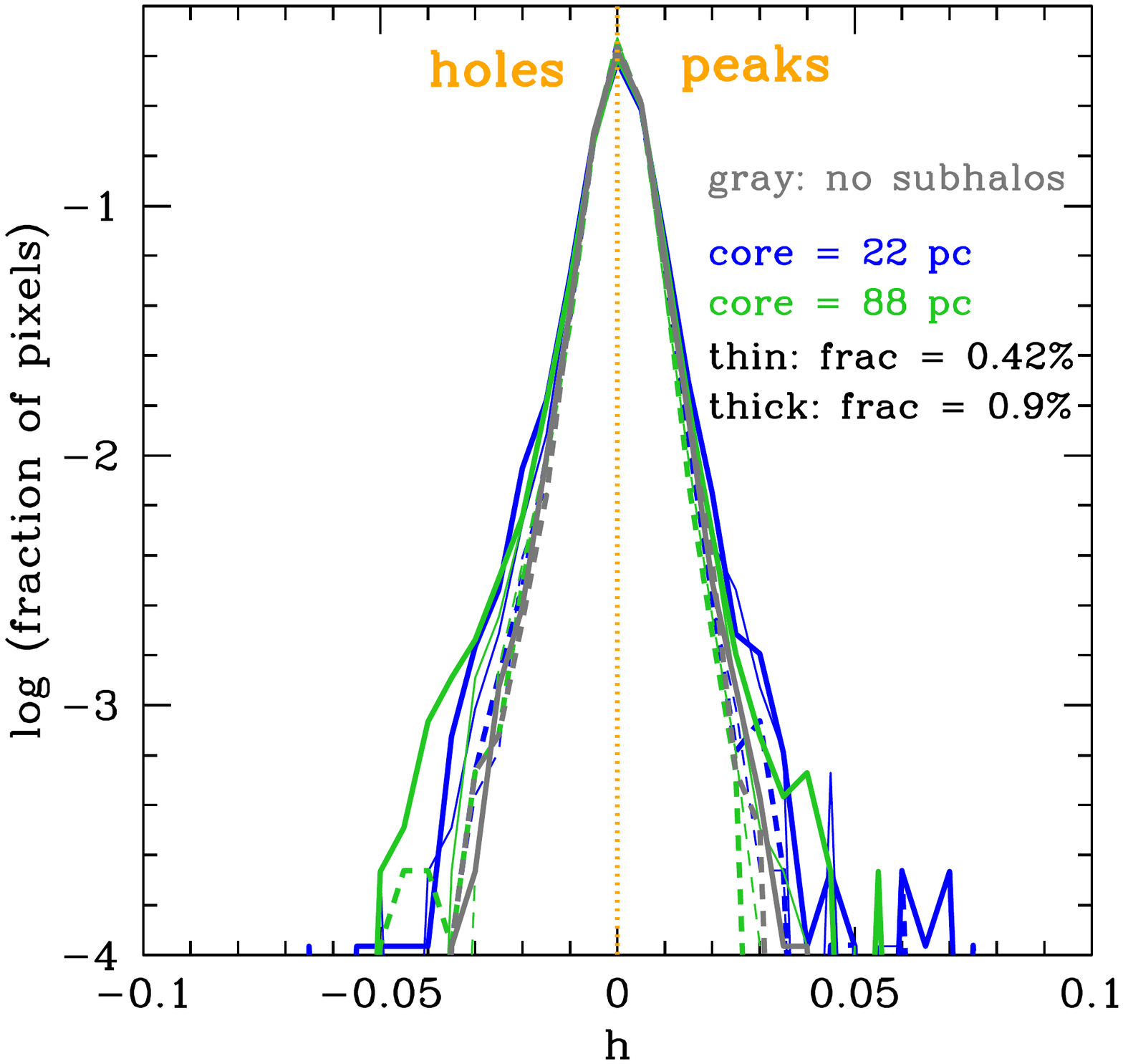}
\includegraphics[trim={0cm 5cm 0cm 4cm},clip,width=0.49\textwidth]{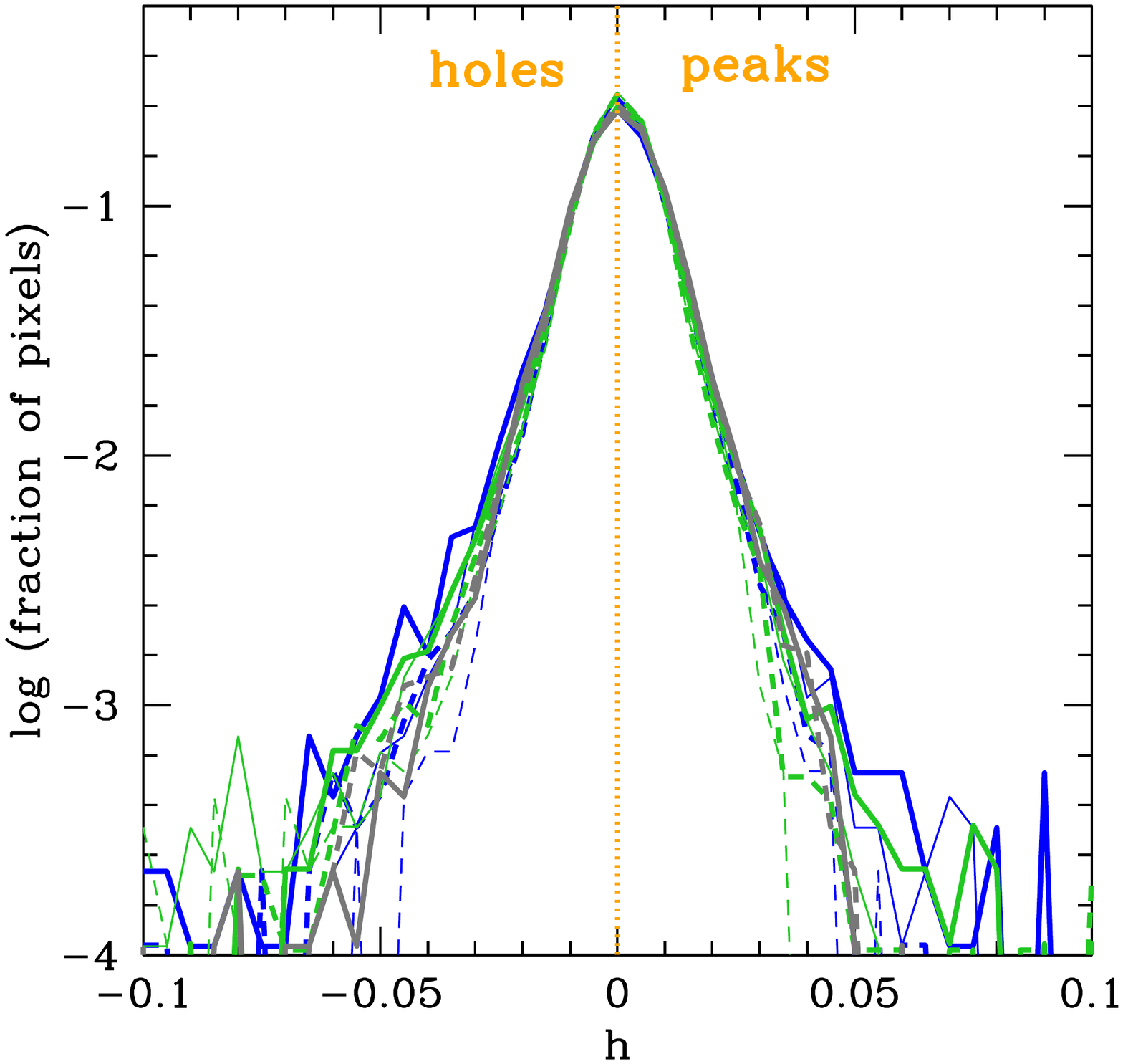}
\caption{Observable signature of subhalos: the fraction of pixels with values of $h$, Eq.~\ref{eq:h}. Positive and negative values of $h$ are for local peaks and holes, respectively. Solid and dashed distributions represent the +ve and $-$ve sides of the cluster CC, respectively. A band of $\pm 0.1''$ around the cluster CC was omitted.
See the legend for model parameters. {\it Left panel:}  assuming that all sources have the same luminosity. {\it Right panel:} sources have a luminosity function $dn/dL\propto L^{-2.5}$. Each line is an average over 16 realizations of the $2\,\square''$ modeling window. \\\\}
\label{fig:diffCCobs}
\end{figure*}

In the left panel of Figure~\ref{fig:diffCCobs} we plot the fraction of $0.04''$ pixels in the lens plane that have $h$ value indicated on the horizontal axis. The holes and peaks are on the left and right sides of each panel, respectively. The model parameters are shown in the left panel. A band of $\pm 0.1''$ surrounding the cluster CC was omitted to isolate the effect of subhalos from that of the cluster CC. The solid and dashed lines represent the +ve and $-$ve sides of the cluster CC, respectively. The gray line shows the subhaloless case. At low $|h|$ values all distributions are very similar. At $|h|\gtrsim 0.03$--0.04, the solid blue and green lines are above the gray lines, meaning there are more pixel-to-pixel variations in flux in the presence of subhalos. More importantly, the solid lines tend to be above the corresponding dashed lines, implying that more of the pixel-to-pixel variations in flux are on the +ve cluster side.

This is the type of signature that is in principle observable. The value $|h|=0.04$ corresponds to a $0.1$\,mag difference in flux. If all sources have the same luminosity, as the left panel assumes, flux peaks and holes are probably only barely measurable given the typical noise level of observations. 

The right panel of Figure~\ref{fig:diffCCobs} is the same as the left, except that the sources were drawn from a luminosity function of the form $dn/dL\propto L^{-2.5}$. Here, the distribution is significantly wider than on the left panel, implying that larger values of $|h|$ are possible; in fact, for $|h|>0.05$ (or 0.125\,mag), there should be very few (fraction $<$ few$\times 10^{-4}$) holes and peaks on the $-$ve cluster side, or in a subhaloless case. On the +ve side of a cluster with subhalo, the fraction of pixels with these value of $|h|$ should be several times higher.
Given these numbers, determining if subhalos are present or absent may be possible using one region of the cluster of size $1''$ to a few arcseconds, but extracting subhalo parameters such as subhalo mass fraction or core sizes will require more than one region in a galaxy cluster, and possibly more than one cluster. 

To sum up, a signature of subhalos is that the number of peaks and holes of progressively higher amplitude gets larger on the +ve side of the cluster compared to the $-$ve side. This asymmetry should be looked for at some distance away from the cluster CC, $\sim \pm 0.1''$, so as not to confuse it with that due to microlensing by stars near the cluster CC.

We note that pixel-level detection relies on the fact that point sources, and hence lensed images, have a finite projected number density on the sky. If the source density were very high, or if the source were a truly continuous (though not necessarily uniform) surface brightness distribution, these tests would not work. They rely on discreteness and stochasticity of image distribution.

\subsection{Individual Highly Magnified Images}\label{sec:indiv}

When subhalos magnify \llrw{sources} by a factor of a few $100$ to $1000$, their flux will contribute to the total flux of the pixel they are in, as discussed above.  But if an image is magnified by more than a few $1000$ it can be detected on its own. This has been the case with individual star detections \cite[e.g.,][]{kel18}. Very high magnifications are possible by subhalos alone, or with the help of microlensing.

\subsubsection{Magnification by Subhalos Alone}

Figure~\ref{fig:imageLF} shows that the high-magnification tail of the image distribution in the lens plane scales as $|\mu|^{-2}$. Extending that relation to higher magnifications suggests that fewer than 1 in $10^6$ images will attain $|\mu|>10^4$ and be detectable as brightened images of individual stars. Since the number of images per $\sim 1''$--$2''$ region is of order a million, there is a reasonable chance of detecting subhalo-brightened images of individual stars. If these happen to lie close to the cluster CC, it will be hard to ascertain that they were brightened by a subhalo, since microlensing is more likely to be responsible. Subhalos are less likely to produce rapid time variability, compared to   microlensing.

\subsubsection{Magnification by Subhalos and Microlensing}\label{sec:micro}

If an image that was moderately magnified by a subhalo also gets microlensed, then the combined magnification can be high enough to make it detectable as an individual highly magnified star. 

To estimate the statistical effect of including microlensing by stars and other point lenses in the cluster, one would convolve the image luminosity function with the microlensing probability distribution. Convolving steep image flux distributions with shallower magnification probability distributions, like that of microlensing, will result in a shallower observed luminosity function. Consulting Figure~\ref{fig:imageLF}, which plots image luminosity function, we see that microlensing will have the largest impact on the images on the $-$ve cluster side and subhalos with larger core radii, shown as the thick dotted green line, because their image flux distribution has a sharp drop off at intermediate $|\mu|$'s. Therefore, convolving with a microlensing probability density function (PDF) will extend these distributions to higher magnifications. This will increase the number of highly magnified images on the $-$ve side, but their number should still remain small compared to the ones on the +ve side.

The PDF of image magnifications from microlensing differs on the +ve and $-$ve sides, as shown in Figure 14 of \cite{die18}: on the $-$ve side, where image demagnifications are usually possible, the PDF has a flat plateau for $|\mu|$ up to $\sim 300$, and a high-magnification tail scaling as $|\mu|^{-3}$. 

\subsection{Where to Look for Subhalos}

The calculations in this paper were tailored to a cluster of mass similar to that of SGAS J1226+2152, which was studied by \cite{dai20} who found variations in flux on either side of the cluster CC. Other cluster candidates include Hamilton's object \citep{gri21}, and Lyman-$\alpha$ detection in RXJ 1347.5-1145 \citep{ric21}.

In general, any images straddling the cluster CC are good targets to look for subhalo lensing, especially not in the immediate vicinity of the cluster CC, because as we saw in Figures~\ref{fig:imageplane} and \ref{fig:countnearCC}, subhalo-magnified images can appear farther away from the cluster CC, where they are less likely to be confused with microlensing-brightened events. Microlensed stars are also more likely to occur in close pairs such as in \cite{kel18} and \cite{che19}, and are more likely to give rise to time variability compared with subhalos. 

\subsection{Subhalos vs. Nonstandard Forms of Dark Matter}

While dark matter is generally assumed to consist of massive nonrelativistic particles, it is possible that its nature is very different. Wave, or fuzzy dark matter, composed of ultralight particles, can introduce both positive and negative mass fluctuations about the mean density throughout the cluster as a result of quantum interference \citep{schive,hui,lar22,amruth}.
These pervasive fluctuations have been shown to produce significant perturbations to the brightness and positions of lensed images in galaxy-scale lenses \citep{amruth}, while also being able to reproduce the observations of specific lensing systems that smooth lens models typically struggle to explain. These results highlight the need for some form of mass substructure in the lens. However, the effect of wave dark matter is expected to be smaller on cluster compared to galaxy scales and remains to be explored in detail, with preliminary results presented by \cite{kel22}. Cold dark matter (CDM) subhalos, on the other hand, lead to an asymmetry between +ve and $-$ve sides of clusters. The presence or absence of asymmetry may be a test for the nature of dark matter.

\subsection{Clusters with Shallower Density Profiles}

So far we worked with the cluster lens whose 3D density profile is that of a singular isothermal sphere. However, shallower profiles are also possible at radii where cluster CCs are formed. In Appendix~\ref{appB} we consider a situation very similar to the one above, but with projected density $\propto r^{-0.9}$. 

Shallower cluster profiles produce notable differences: highly magnified images extend farther away from the cluster CC (Figure~\ref{fig:appBboth1}), and should therefore be looked for farther from the cluster CC. The flux peaks and holes are more strongly correlated with subhalos (Figure~\ref{fig:appBboth2}, right panel), and the amplitudes of flux peaks and holes, $|h|$, are larger (Figure~\ref{fig:diffCCobsDIR82}) and will therefore be more easily detected than in clusters with steeper density profiles.

\section{Summary and Conclusions}

In this paper we investigated the effect of dark matter subhalos on lensing near the tangential critical curves of galaxy clusters. We started by first calculating the properties (distribution and magnification) of lensed images highly magnified by subhalos. We then used analytical and numerical experiments to examine and explain the various effects of subhalo lensing. In the last portion of the paper we explored whether subhalos have detectable signatures. We divide these into two categories: pixel-level signatures, where subhalo magnifications contribute toward variations in flux between neighboring detector pixels, and images that are so highly magnified that they can be detected on their own as individual stars in source galaxies. In the latter case high magnification can be due to subhalos alone, or aided by microlensing. Our work is preliminary; the actual analysis will need to take into account noise and its properties, which can be nontrivial, arising not only from noise in empty fields, but from intracluster light as well.

Just like microlensing by intracluster stars, subhalos can magnify stars in background galaxies by factors of hundreds to thousands. In both cases, high magnifications happen preferentially on the positive parity side of the cluster. However, there are some important differences between subhalo lensing and stellar microlensing:

$\bullet$ Because subhalos are generally significantly more massive than microlenses,  their influence extends farther away from the cluster critical curve (Figures~\ref{fig:imageplane}, \ref{fig:countnearCC}, and \ref{fig:appBboth1}).  
If magnified images are found far from the cluster CC, they are more likely to have been magnified by subhalos rather than by microlensing \citep{mee22a}.

$\bullet$ If the cluster density profile is a power law (instead of linear), subhalos of a given mass will produce longer critical curves around themselves on the positive parity side, compared to the negative side. This can lead to a significant asymmetry on the two sides of the cluster: there will be more highly magnified images on the positive parity side (Figures~\ref{fig:imageplane}, \ref{fig:countnearCC}, and \ref{fig:appBboth1}). The asymmetry is stronger for steeper cluster density slopes, and is more pronounced for subhalos than microlenses because the former's effect extends farther away from the cluster critical curve (Section~\ref{sec:gradient}).

$\bullet$ The number of highly magnified images is proportional to the length of the subhalo critical curves, with approximately the same linear relation applicable to both the positive and negative parity sides of the cluster (Figures~\ref{fig:subhaloCC} and \ref{fig:appBboth2}, left panel).

$\bullet$ The fact that subhalos have an extended mass distribution, possibly with a flat density core (Section~\ref{sec:reasons3}), increases the asymmetry between the image properties on the two parity sides of the cluster, making it more likely that bright images will appear on the positive side (Section~\ref{sec:reasons}).
Because of their density cores, the asymmetry is stronger compared to the case with microlenses, and can help differentiate between microlens- and subhalo-magnified images, as well as between small and large subhalo core radii.

$\bullet$ Subhalos with larger core radii generate shorter critical curves around themselves (see Figure~\ref{fig:DEMOimageplane} \llrw{for an example of subhalos on the positive parity cluster side}), resulting in fewer highly magnified images. \llrw{The most dramatic case is for core radii $\gtrsim 50\,$pc and subhalos on the negative parity sides; in these cases the critical curves around subhalos disappear altogether, and so subhalos produce no highly magnified images (Figure~\ref{fig:imageplane} and \ref{fig:imageLF})}.

$\bullet$ Another consequence of subhalo mass being significantly larger than that of microlenses is that their lensing effects are not transient or time variable, unless subhalo lensing is augmented by stellar microlensing, \new{or if the subhalo has low mass, $\lesssim 10^6\,M_\odot$, and the velocity vector of the subhalo relative to the source star is well aligned with the image stretching due to the main cluster.}

\begin{acknowledgments}
L.L.R.W., P.L.K., T.T., and A.V.F. would like to acknowledge HST programs GO-15936 and GO-16278, and SNAP program GO-16729; \llrw{financial support was provided by the Space Telescope Science Institute, which is operated by the Association of Universities for Research in Astronomy, Inc., under NASA contract NAS5-26555.}
P.L.K. acknowledges NSF AST-1908823 \llrw{and AST-2308051} grants.
J.M.D. acknowledges the support of projects PGC2018-101814-B-100 and MDM-2017-0765. 
\llrw{A.V.F. was supported by the Christopher R. Redlich Fund and numerous individual donors.}
A.K.M. and A.Z. acknowledge support by grant 2020750 from the United States-Israel Binational Science Foundation (BSF) and grant 2109066 from the United States National Science Foundation (NSF), and by the Ministry of Science \& Technology, Israel. 
\end{acknowledgments}

\appendix

\section{Subhalo 2D Projected Profiles}\label{appA}

We would like to use subhalos whose lensing properties are simple analytical functions. At the same time, their profiles have to closely resemble profiles obtained in numerical dark matter simulations, i.e., NFW profiles, but with a sharper dropoff in density at large radii. For simplicity, we deal with circularly symmetric profiles. \new{This is a good approximation because most of the elongation in the problem is provided by the main cluster.}

We start with specifying the deflection angle. We want it to increase at small radii  (similar to that of a flat density core), then reach a maximum, and then decrease at large radii, approximating the behavior of steeply declining density profile. We chose the function 
\begin{equation}
    |\vec\alpha|=\alpha=[(a/r+1)(r/b+1)]^{-1}\, , \label{eq:alpha}
\end{equation}
where $r$ and $a$ are dimensionless radii, and $a$ is a constant.
The first term dominates at small radii, the second at large radii. \new{The first term mimics that of a softened isothermal sphere, while the second falls like that of a point mass.}
The dimensionless mass enclosed within $r$ is
\begin{equation}
    m(r)=\frac{r^2}{r^2/b+r(1+a/b)+a}\, . \label{eq:menc}
\end{equation}
The projected surface mass density, in units of critical density, is
\begin{equation}
    \kappa(r)=\frac{\frac{1}{2}r(1+a/b)+a}{[{r^2/b+r(1+a/b)+a}]^2}\, . \label{eq:kappa}
\end{equation}
The lensing potential is given by
\begin{equation}
    \psi(r)=\frac{b}{(a-b)}\Big(a\,\ln[a+r]-b\ln[b+r]\Big)\, . \label{eq:pot}
\end{equation}
One can show that $\vec\nabla\psi(r)=\vec\alpha(r)$, and that the second derivatives of $\psi$ can be combined through the lensing Poisson equation to give $\kappa(r)$: $\frac{1}{2}(\psi_{,xx}+\psi_{,yy})=\kappa$.

\begin{figure}
    \centering
    \includegraphics[trim={0.5cm 2cm 0.5cm 5cm},clip,width=0.65\textwidth]{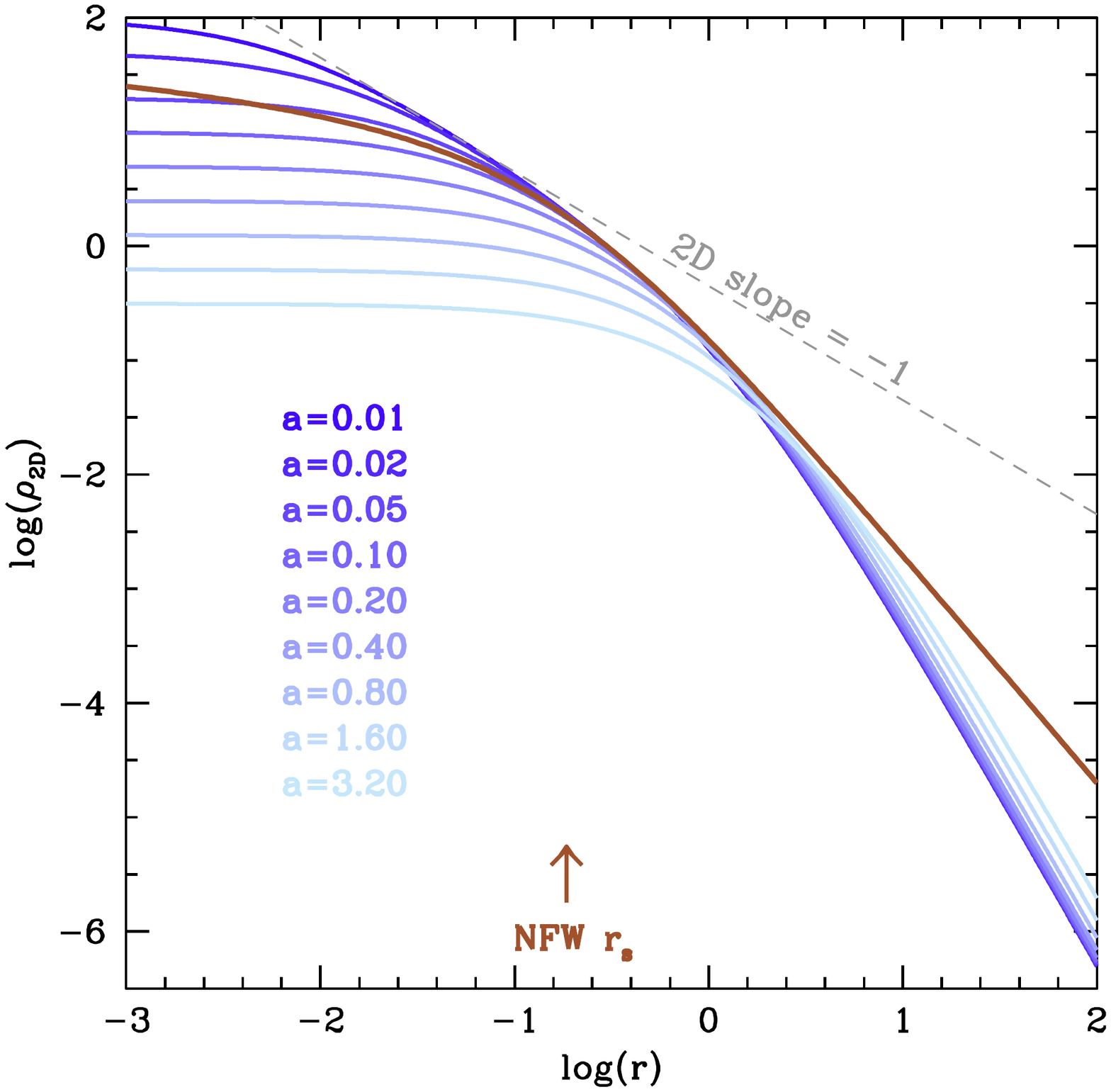}
    \vskip-2cm
    \caption{The function we use to represent subhalos is shown as blue curves. The parameter $a$ determines the inner profile shape. In this paper we use $a=0.05$, which closely resembles NFW (brown curve), except for large radii where our function falls off quicker, approximating tidal truncation. The 2D slope of $d\ln\rho_{2D}/dr=-1$, which corresponds to an ''isothermal'' slope in 3D, is shown as a dashed gray line. } 
    \label{fig:myy}
\end{figure}

The 2D density profiles for a range of $a$ values are shown in Figure~\ref{fig:myy}. A range of shapes can be obtained by varying $a$. Smaller $a$ result in smaller, flattish density cores and a gentler transition to steeper outer profiles. On the other hand, larger $a$ have extended, flat density cores, and a sharper transition to the outer steep slope. The outer profile always goes as $r^{-3}$ in projection, ensuring that mass converges quickly. 

For comparison, we plot the NFW profile in brown. It has been rescaled vertically and in radius to show the close resemblance to one of our functions: third from the top, with $a=0.05$. The small vertical arrow at the bottom of the plot shows the NFW's scale radius $r_s$. NFW declines more slowly at larger radii, as $r^{-2}$ in projection.

\section{Subhalos in a Cluster with a Shallower Density Profile}\label{appB}

\begin{figure*}[!h]
    \centering
    \includegraphics[trim={0cm 5cm 0cm 4cm},clip,width=0.49\textwidth]{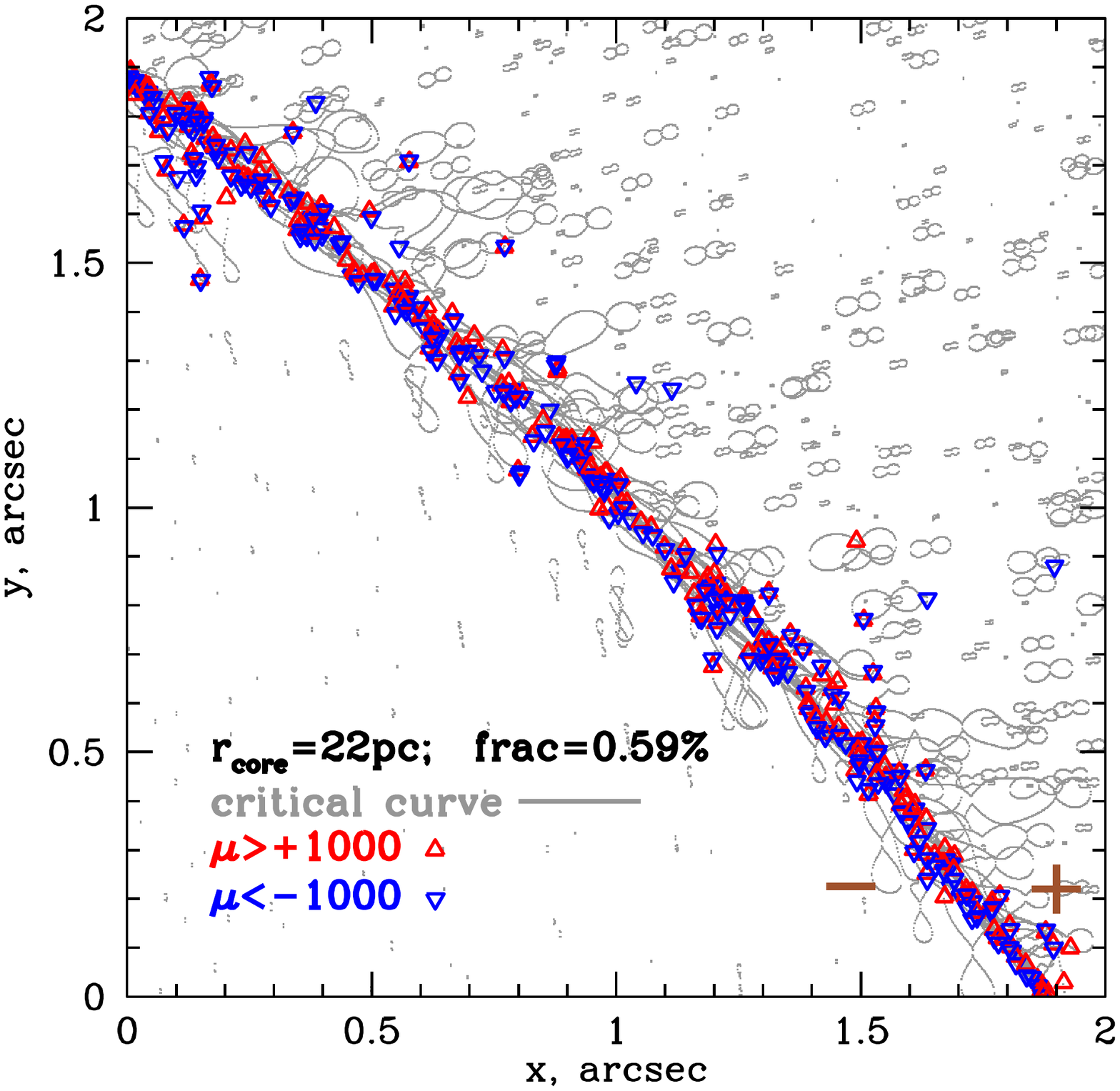}    
    \includegraphics[trim={0cm 5cm 0cm 4cm},clip,width=0.49\textwidth]{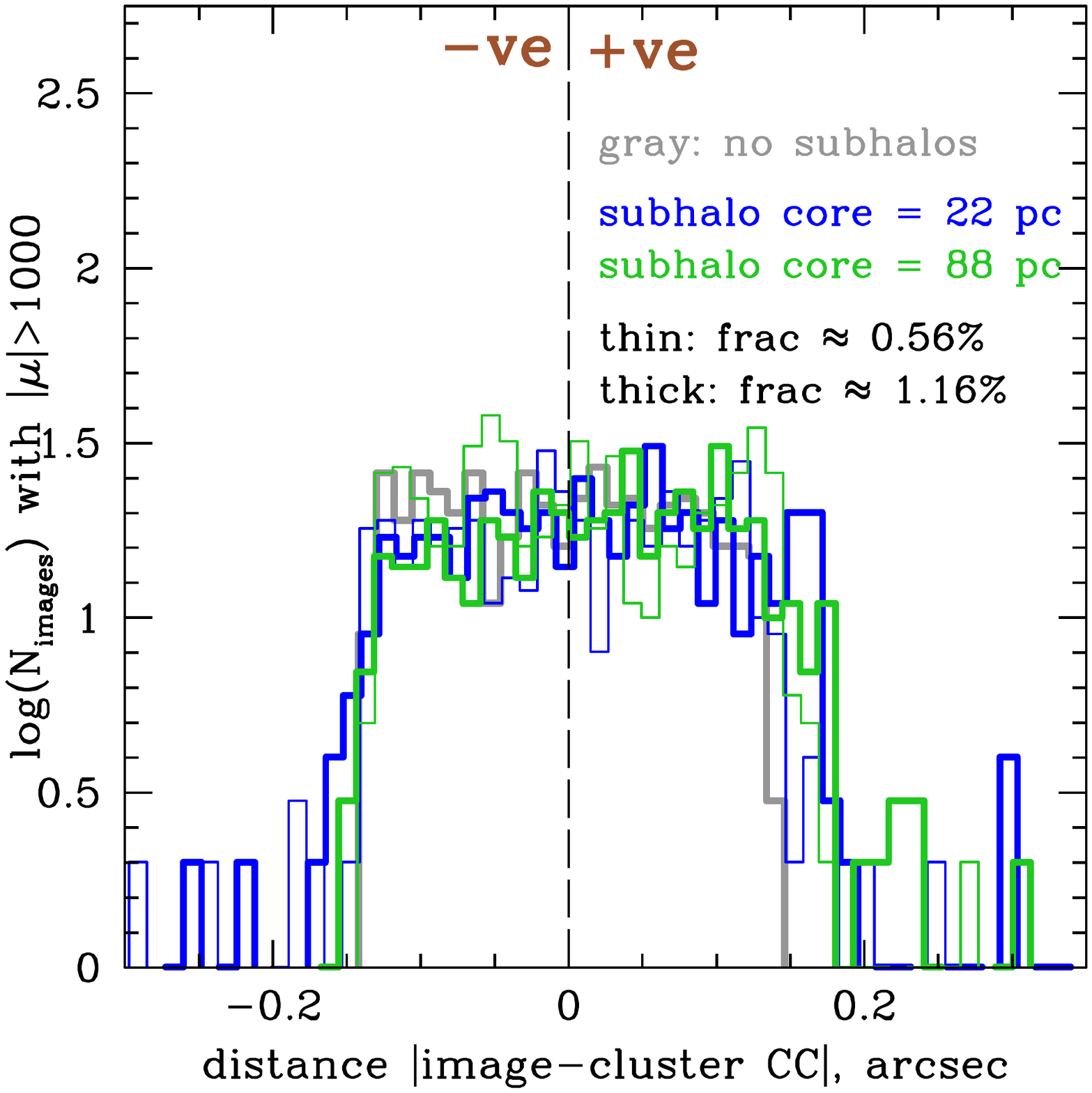}    
    \caption{
    {\it {Left:}} Similar to Figure~\ref{fig:imageplane}, but for a shallower cluster density slope.
    {\it Right: } Similar to Figure~\ref{fig:countnearCC}, but for a shallower cluster density slope. Note that the horizontal axis is wider than in the former figure.}
    \label{fig:appBboth1}
\end{figure*}

\begin{figure*}[!h]
    \centering
    \includegraphics[trim={0cm 5cm 0cm 4cm},clip,width=0.49\textwidth]{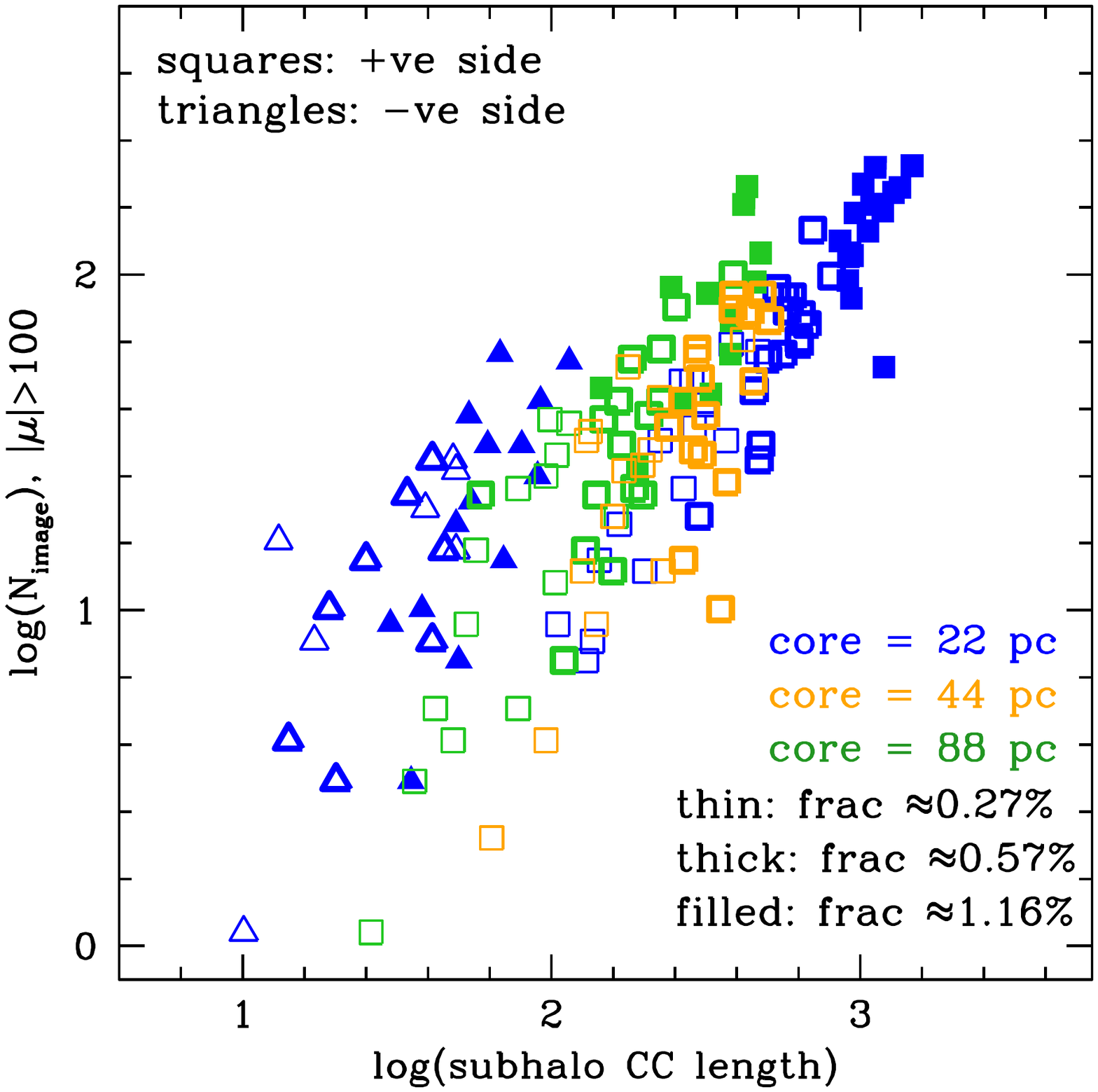}    
    \includegraphics[trim={0cm 5cm 0cm 4cm},clip,width=0.49\textwidth]{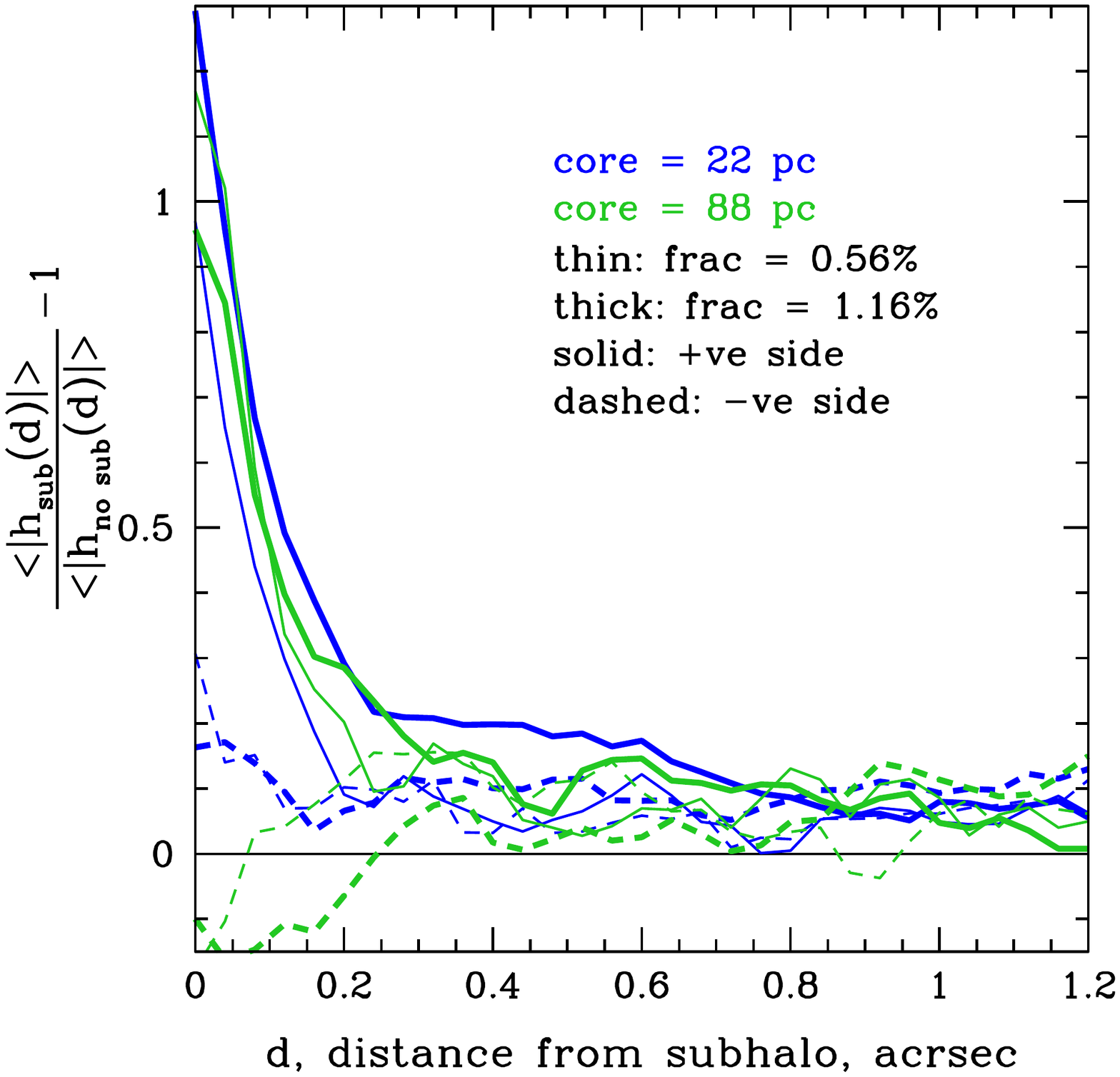}
    \caption{{\it Left: } Similar to Figure~\ref{fig:subhaloCC}, but for a shallower density slope of the cluster. 
    {\it Right:} Similar to Figure~\ref{fig:correl}, but for a shallower density slope of the cluster. Note that the vertical axis goes to higher values than in the former figure.}
    \label{fig:appBboth2}
\end{figure*}

\begin{figure*}
\includegraphics[trim={0cm 5cm 0cm 4cm},clip,width=0.49\textwidth]{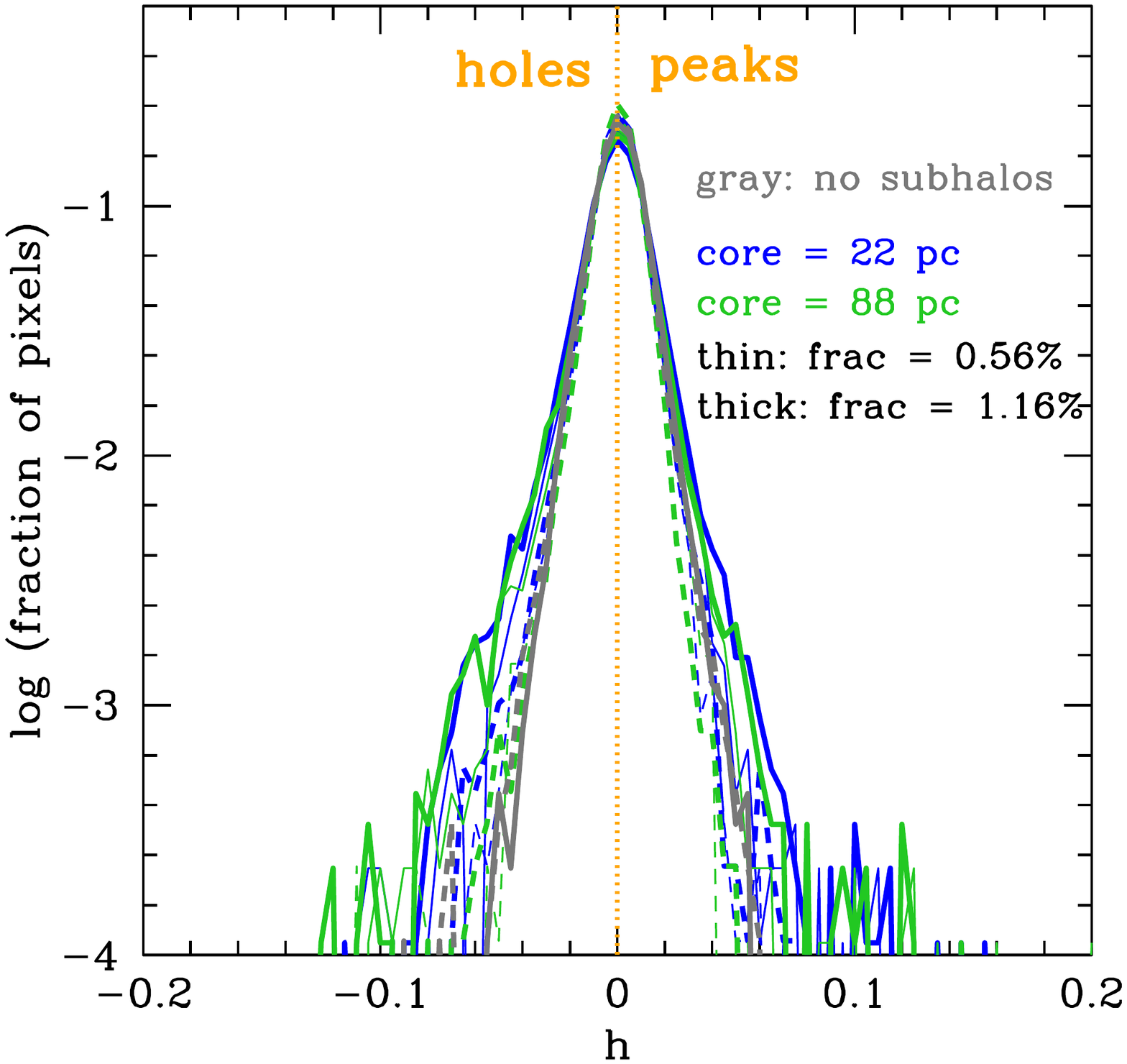}
\includegraphics[trim={0cm 5cm 0cm 4cm},clip,width=0.49\textwidth]{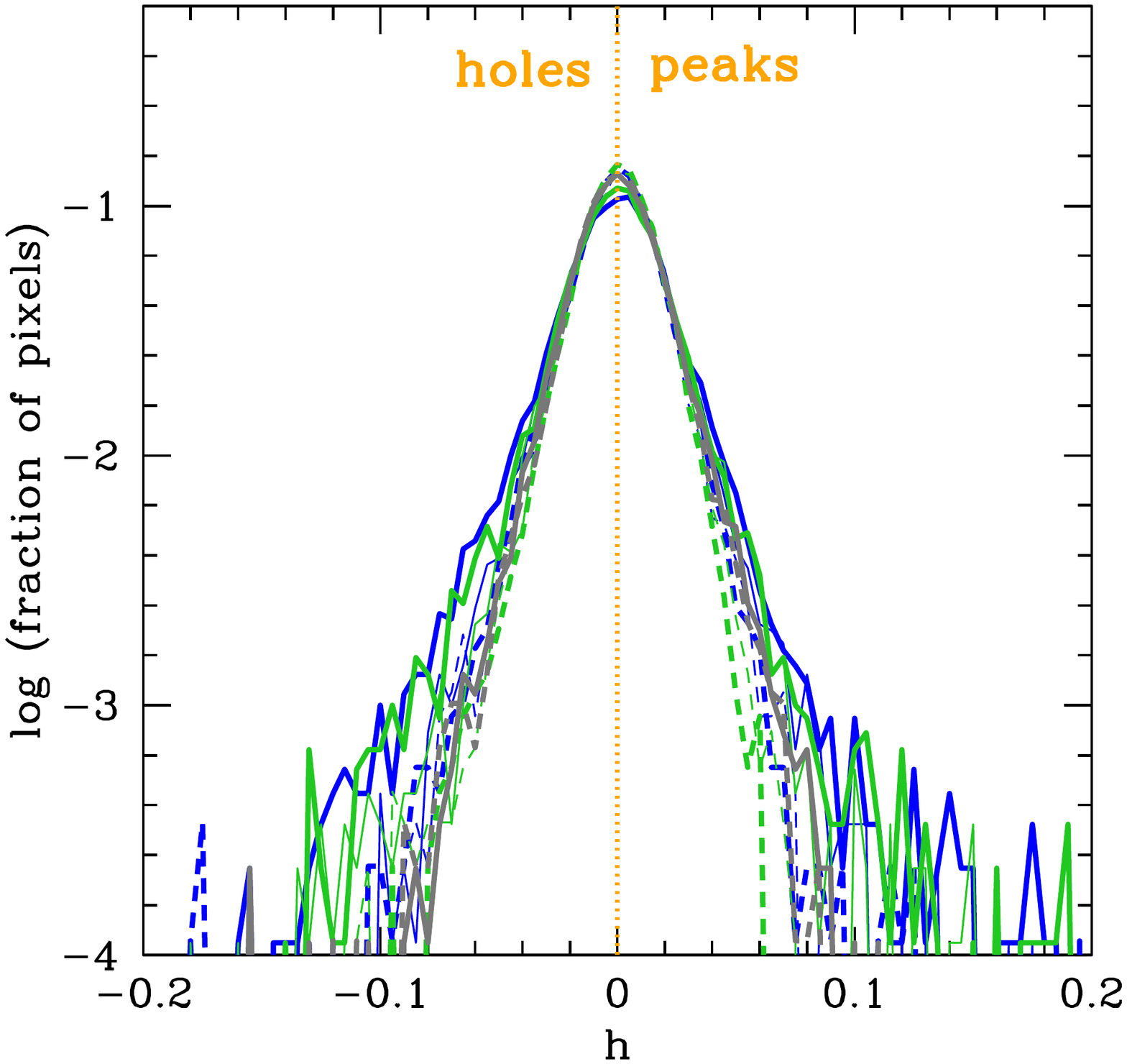}
\caption{Similar to Figure~\ref{fig:diffCCobs}, but for a shallower density slope of the cluster. Note that the horizontal axis in this figure is twice as wide as in the former.}
\label{fig:diffCCobsDIR82}
\end{figure*}

In the main paper we considered a cluster whose density profile slope is that of an isothermal sphere, $\propto {r_{2D}}^{-1}$ in projection. One of our findings is that the cluster density slope is an important factor in determining the properties of highly magnified images formed near its critical curve. For that reason, in this Appendix we consider a somewhat shallower projected density profile that is also consistent with observations of clusters, $\propto {r_{2D}}^{-0.9}$. The plots presented here (Figures~\ref{fig:appBboth1}, left and right panels; \ref{fig:appBboth2}, left and right panels; and Figure~\ref{fig:diffCCobsDIR82}) are similar to the corresponding ones in the main text (Figures~\ref{fig:imageplane}, \ref{fig:countnearCC}, \ref{fig:correl}, and \ref{fig:diffCCobs}). 

The conversions relating linear and angular scales in the source and lens plane are the same as in the main text, but the number density of points sources in the source plane is 0.03\,stars\,pc$^{-2}$ here. As in the main paper, we are concerned only with intrinsically luminous stars, which are rare, motivating the low number density.

The distribution of images around the cluster CC here is wider than for the steeper cluster profile; in fact, the right panel of Figure~\ref{fig:appBboth1} shows that the distribution does not peak at the cluster CC; the images are uniformly distributed within $\sim \pm 0.15''$. \new{This flattening of the distribution is the direct result of the shallower cluster density profile, which makes the region where $(1-\kappa-\gamma)\approx 0$ into a wider band around the CC. It is similar to the case of microlenses that reach optical depth $\sim 1$ near the cluster CC. The same effect is also seen in models with wave dark matter, where the key parameter is not cluster slope or microlensing optical depth, but the mass of the boson.} The asymmetry seen in the steeper cluster density case, where bright images were preferentially found on the +ve cluster side, is also present here, but is seen farther away from the cluster CC, at $\gtrsim 0.15''$. The spatial correlation of local flux peaks and holes with subhalo locations is stronger here (right panel of Figure~\ref{fig:appBboth2}) compared to the case of a steeper cluster profile. That means that flux peaks and holes are more likely to indicate the locations of subhalos. The amplitude of local flux peaks and holes (i.e., values of $|h|$) are also larger here (Figure~\ref{fig:diffCCobsDIR82}).

\section{Why the Subhalo CC's Look Like ``8'' and ``$\infty$''}\label{appC}

The gray lines in the left panel of Figure~\ref{fig:dir51} are tangential critical curves \citep[see also][]{cha84,ogu18,die18}. In the absence of subhalos, the line would have been an uninterrupted smooth approximate diagonal. The two subhalos create additional loops. The condition for the tangential critical curve in the presence of subhalos can be written as
\begin{equation}
   0=1-(\kappa_m+\kappa_s)-([\gamma_{1m}+\gamma_{1s}]^2+[\gamma_{2m}+\gamma_{2s}]^2)^{1/2}\, . 
   \label{eq:tanCCexact}
\end{equation}
Here, subscripts $m$ and $s$ stand for the {\it m}ain cluster and
{\it s}ubhalo, respectively.

\begin{figure}
    \centering
    \includegraphics[trim={0cm 14cm 0cm 3.5cm},clip,width=0.495\textwidth]{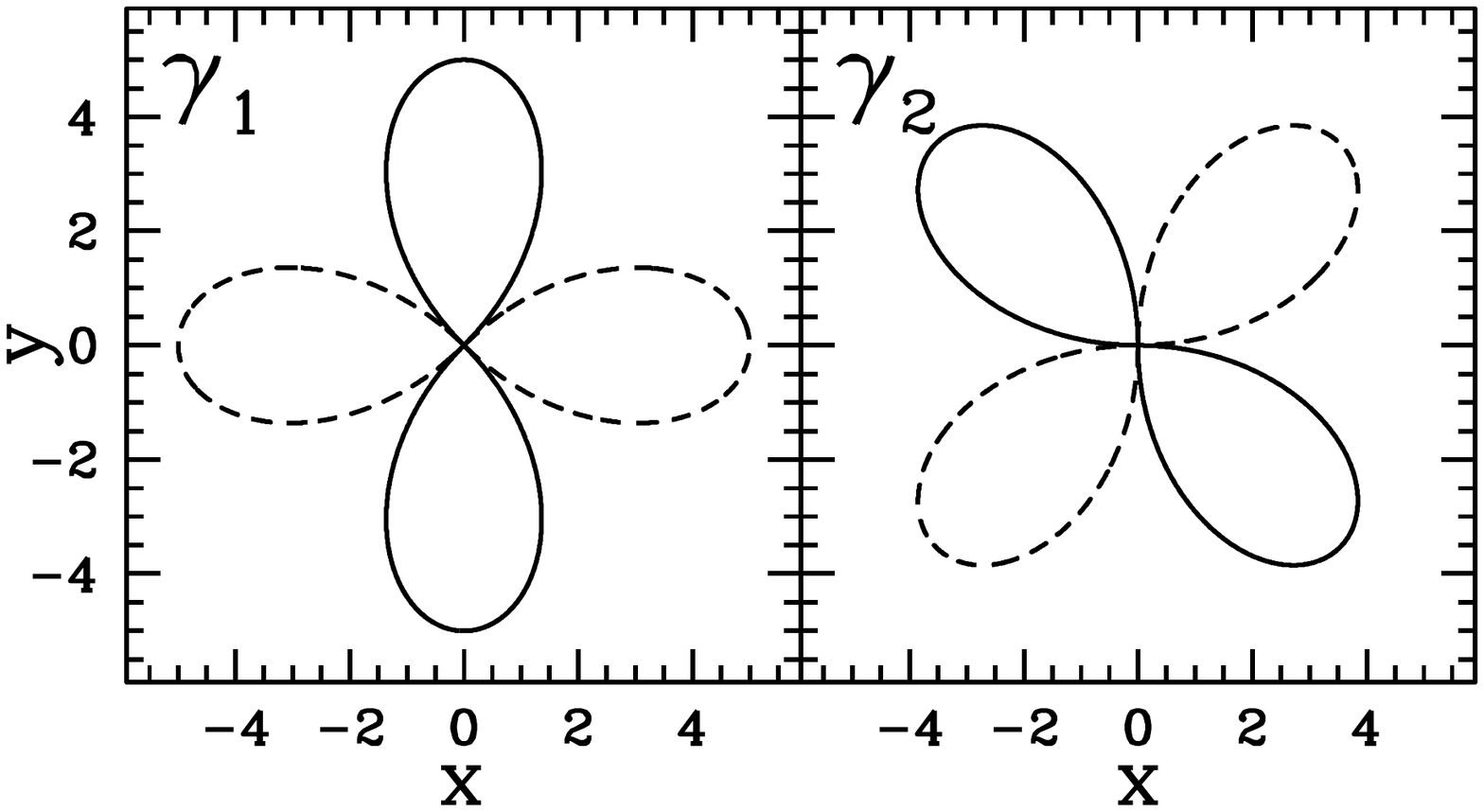} 
    \caption{Contours of equal shear ($\gamma=0.1$) of a SIS lens, Eq.~\ref{eqn:params}; {\it left: }$\gamma_1$; {\it right:} $\gamma_2$. Positive and negative values are shown as solid and dashed lines, respectively.  Location (0,0) is the lens center. Note that even for a non-SIS density profile, like that of a subhalo, this four-leaf structure with alternating positive and negative values is approximately the same. We use this structure in Section~\ref{sec:reasons} to illustrate the values of $\gamma_{1m}$ and $\gamma_{1s}$, and to show that $\gamma_{2m}\approx 0$ and $\gamma_{2s}\approx0$, since our modeling frame is along the positive $x$ axis, where $y=0$.\\}
    \label{fig:gammas}
\end{figure}

Figure~\ref{fig:gammas} shows the pattern of $\gamma$'s for the SIS lens. The center is at the center of the cluster, and our modeling window is on the positive $x$ axis, where $y=0$ (see Figure~\ref{fig:images00}), which means that $\gamma_{2m}\approx 0$. Figure~\ref{fig:gammas} can also be used to approximately estimate the $\gamma$ values of subhalos. In that case the $(0,0)$ in the figure would correspond to the center of a subhalo. The subhalo density profile is not SIS, but the general four-leaf pattern of the $\gamma$ contours with alternating positive and negative values is the same. That means that for subhalos $\gamma_{2s}\approx 0$. 

The density profile of any realistic subhalo in a cluster will not be a SIS. If its profile shape is similar to those from simulations, its central density slope will be somewhat shallower than ${r_{3D}}^{-2}$ (or ${r_{2D}}^{-1}$), and considerably steeper than that at the outskirts, owing to tidal truncation. For such a profile, $\kappa_s>\gamma_s$ in the central region (in the limiting case of a very large flat density core, $\kappa\gg\gamma$), but farther out, $\kappa_s<\gamma_s$. (As a limiting case of a centrally concentrated lens, consider a point-mass lens where $\kappa=0$ and so is always smaller than $\gamma$, outside of the actual point.)  Let us consider such a centrally concentrated subhalo.
At the location of subhalo CC, which is some distance away from its center, its surface density is small compared to its shear, and we can assume that $\kappa_s\approx 0$. 
The tangential CC condition of Eq.~\ref{eq:tanCCexact} is now reduced to 
\begin{equation}
    0\approx 1-\kappa_m-|\gamma_{1m}+\gamma_{1s}|\, .
    \label{eq:tanCCapprox}
\end{equation}

Next, we use Figure~\ref{fig:gammas} to estimate $\gamma$ values for the SIS cluster lens. The center is at the center of the cluster, and our modeling window is on the positive $x$ axis ($y=0$), and is small compared to the extent of the cluster, which means that $\gamma_{1m}$ is negative everywhere in our modeling frame; since we already established that $\gamma_{2m}\approx0$, we conclude that $\gamma_{1m}\approx -\kappa_m$.

The only unknown in Eq.~\ref{eq:tanCCapprox} is $\gamma_{1s}$. Because the other two terms in that equation refer to the cluster, their values will stay approximately constant near the subhalo CC. Therefore, the shape of the subhalo CC will follow the shape of $\gamma_{1s}\approx\,$const, i.e., one of the shapes in Figure~\ref{fig:gammas}. To determine which one, we need to figure out the sign of $\gamma_{1s}$ for the two subhalos on the two sides of the cluster CC.

Let us first consider the case of $\gamma_{1s}>0$. Equation~\ref{eq:tanCCapprox} can be rewritten as $0=1-2\kappa_m+\gamma_{1s}.$ This can be satisfied only if $\kappa_m>\kappa_{\rm CC}$, where $\kappa_{\rm CC}$ is at the location of the cluster CC, and is 0.5 for a SIS. The cluster density is larger than $\kappa_{\rm CC}$ on the $-$ve side. The shape of $\gamma_{1s}$ when its value is positive is an upright ``8''; therefore, subhalo CCs will have that shape on the $-$ve cluster side. This is what we see in Figure~\ref{fig:dir51}.

Now let us consider the case of $\gamma_{1s}<0$. Equation~\ref{eq:tanCCapprox} can be rewritten as $0=1-2\kappa_m-|\gamma_{1s}|.$ This can be satisfied only if $\kappa_m<\kappa_{\rm CC}$. The cluster density is lower than $\kappa_{\rm CC}$ on the +ve side. The shape of $\gamma_{1s}$ when its value is negative is an ``$\infty$" sign; therefore, subhalo CCs will have that shape on the +ve cluster side. This is what we see in Figure~\ref{fig:dir51}.


\end{document}